\documentclass[format=acmsmall, review=false, screen=true]{acmart}

\renewcommand\footnotetextcopyrightpermission[1]{} 
\usepackage[utf8]{inputenc}

\usepackage{booktabs} 

\usepackage[ruled]{algorithm2e}

\SetAlFnt{\small}
\SetAlCapFnt{\small}
\SetAlCapNameFnt{\small}
\SetAlCapHSkip{0pt}
\IncMargin{-\parindent}

\usepackage[group-separator={}]{siunitx}

\usepackage{caption,subcaption}

\usepackage{mathtools}
\DeclarePairedDelimiter\ceil{\lceil}{\rceil}

\usepackage{multirow}
\usepackage{array}
\usepackage{booktabs}
\newcolumntype{P}[1]{>{\centering\arraybackslash}p{#1}}
\newcolumntype{M}[1]{>{\centering\arraybackslash}m{#1}}

\usepackage{float}

\newcommand{\MAN}{\textsf{MAN}\xspace}

\setcopyright{acmcopyright}
\acmJournal{JETC}
\acmYear{2019} \acmVolume{1} \acmNumber{1} \acmArticle{1} \acmMonth{1} \acmPrice{15.00}\acmDOI{10.1145/3314326}

\begin{document}
\title[Hardware Optimizations of Dense Binary Hyperdimensional Computing]{Hardware Optimizations of Dense Binary Hyperdimensional Computing: Rematerialization of Hypervectors, Binarized Bundling, and Combinational Associative Memory}



\author{Manuel Schmuck}
\affiliation{%
	\institution{ETH Zürich}
	\city{Zürich}
	\country{Switzerland}}
\email{schmucma@student.ethz.ch}

\author{Luca Benini}
\affiliation{%
	\institution{ETH Zürich}
	\city{Zürich}
	\country{Switzerland}}
\affiliation{%
	\institution{Università di Bologna}
	\city{Bologna}
	\country{Italy}}
\email{lbenini@iis.ee.ethz.ch}

\author{Abbas Rahimi}
\orcid{1234-5678-9012-3456} 
\affiliation{%
	\institution{ETH Zürich}
	\city{Zürich}
	\country{Switzerland}}
\email{abbas@iis.ee.ethz.ch}


\begin{abstract}
%

Brain-inspired hyperdimensional (HD) computing models neural activity patterns of the very size of the brain's circuits with points of a hyperdimensional space, that is, with \emph{hypervectors}.  
Hypervectors are $D$-dimensional (pseudo)random vectors with independent and identically distributed (i.i.d.) components constituting ultra-wide holographic words: $D=10,000$ bits, for instance.
At its very core, HD computing manipulates a set of seed hypervectors to build composite hypervectors representing objects of interest. 
It demands memory optimizations with simple operations for an efficient hardware realization.
In this paper, we propose hardware techniques for optimizations of HD computing, in a synthesizable open-source VHDL library, to enable co-located implementation of both learning and classification tasks on only a small portion of Xilinx\textsuperscript{\textregistered} UltraScale\textsuperscript{\texttrademark} FPGAs:
(1) We propose simple logical operations to \emph{rematerialize} the hypervectors on the fly rather than loading them from memory. 
These operations massively reduce the memory footprint by directly computing the composite hypervectors whose individual seed hypervectors do not need to be stored in memory.
(2) Bundling a series of hypervectors over time requires a multibit counter per every hypervector component.
We instead propose a binarized \emph{back-to-back} bundling without requiring any counters.  
This truly enables on-chip learning with minimal resources as every hypervector component remains binary over the course of training to avoid otherwise multibit components.
(3) For every classification event, an associative memory is in charge of finding the closest match between a set of learned hypervectors and a query hypervector by using a distance metric.
This operator is proportional to hypervector dimension ($D$), and hence may take $\mathcal{O}(D)$ cycles per classification event.
Accordingly, we significantly improve the throughput of classification by proposing associative memories that steadily reduce the latency of classification to the extreme of a single cycle.
(4) We perform a design space exploration incorporating the proposed techniques on FPGAs for a wearable biosignal processing application as a case study.
Our techniques achieve up to $2.39\times$ area saving, or $2337\times$ throughput improvement.
%
%
The Pareto optimal HD architecture is mapped on only $18340$ configurable logic blocks (CLBs) to learn and classify five hand gestures using four electromyography sensors.

\end{abstract}

%
%




\maketitle
\thispagestyle{empty}


\section{Introduction}
%
%
%
%

Hyperdimensional (HD) computing~\cite{HD09,10K14} is a brain-inspired computational approach based on the understanding that brains compute with patterns of neural activity that are not readily associated with scalar numbers.
In fact, the brain's ability to calculate with numbers is feeble.
However, due to the very size of the brain's circuits, we can model neural activity patterns with points of a hyperdimensional space, that is, with \emph{hypervectors}.
When the dimensionality is in the thousands, operations with hypervectors create a computational behavior with remarkable properties~\cite{SDM}.
HD computing builds upon a well-defined set of highly parallel operations with random hypervectors, is extremely robust in the presence of failures, and offers a complete computational paradigm that is easily applied to many learning applications~\cite{TCAS17}.
Examples include 
analogy-based reasoning~\cite{Dollar-Mexico}, 
latent semantic analysis~\cite{Randomindexing}, 
language recognition~\cite{Aditya,Rahimi2016}, 
text classification~\cite{Rasti2016}, 
speech recognition~\cite{Rasanen2015con,VoiceHD_ICRC17}, 
physical activity prediction~\cite{Rasanen14,Rasanen15},
robot learning by demonstration~\cite{Robot_LearnBehHierarchies,Robot_LearnByDem}, and several biosignal processing tasks such as electromyography (EMG)~\cite{EMG-HD,ISCAS18,DAC18,TNNLS18}, electroencephalography (EEG)~\cite{BICT17,MONET17}, electrocorticography (ECoG)~\cite{BioCAS18}, and in general ExG~\cite{Proc18}.

In contrast with traditional approaches, learning in HD computing is fast and computationally balanced with respect to classification by reusing the same algorithmic and architectural constructs for both modes of operation. 
Its learning is not iterative and thus requires far fewer operations than other approaches (see~\cite{Proc18} for an overview).
Another advantage of HD computing is the simplicity of its basic operations, which is an important factor for energy efficiency.
For instance, HD computing achieves 2$\times$ lower energy consumption at iso-accuracy when compared to a highly-optimized support vector machine (SVM) with fixed-point operations on a commercial embedded ARM Cortex M4 processor for an EMG classification task~\cite{DAC18}.
In what follows, we brief these basic operations.



At its very core, HD computing is all about generating, manipulating, and comparing hypervectors as ultra-wide words.
As the first step, $D$-dimensional hypervectors are initially generated with independent and identically distributed (i.i.d.) components.
Second, these \emph{seed} hypervectors are manipulated to construct composite hypervectors, as richer representations, with componentwise arithmetic operations by needing to communicate with only a local component or an immediate neighbor.  
By using dense binary codes for hypervectors~\cite{BSC96}, the arithmetic operations simply involve bitwise XOR, shift (or rotate), and majority gates~\cite{Rahimi2016}.  
Finally, the constructed hypervectors are compared for \emph{similarity} using a distance metric whose computation involves a reduction operator proportional to the hypervector dimension ($D$)~\cite{Rahimi2016,HAM_HPCA17}.
See Section~\ref{sec:background} for details.
%
%
These operators---at the basis of both learning and inference (Section~\ref{sec:biosignals_with_HD})---demand a memory-centric architecture for efficient and local ultra-wide word processing. 
Emerging nanotechnologies with dense 3D integration can provide a natural fit~\cite{ISSCC18,JSSC18,IEDM_2016}.

%

In this paper, we propose hardware techniques to optimize the aforementioned operators to build an efficient acceleration engine on an FPGA.
As a result of our hardware optimizations (Section~\ref{chap:optimisations}), we provide a synthesizable VHDL library\footnote{The library is open-source and available at: \href{https://github.com/eardbi/hd-vhdl-library}{github.com/eardbi/hd-vhdl-library}} of fully configurable modules exploring trade-offs between area and throughput of the operators.
Our contributions are as follows:

(1) We propose a generic hypervector manipulator (\MAN) module as a fully combinational logic consisting of OR-XOR gates and preprogrammed connections.
The \MAN module substitutes the expensive memory storage for maintaining seed hypervectors with cheaper logical operations to \emph{rematerialize} them.
Hence, representations of composite hypervectors are constructed directly by rematerializing the seed hypervectors as a consequence of reusing the generic \MAN modules that form a combinational network architecture without requiring any memory storage.
%
%
%
%

(2) The arithmetic operations of HD computing with dense binary code exhibit their simplest form by performing local and bitwise operations on binary components.
This however does not hold for the majority gate when it is applied to bundle a series of hypervectors over time, i.e., among different training examples. 
Implementation of the majority gate requires to maintain intermediate (i.e., partially bundled) hypervector representation using a set of $D$ multibit counters---every counter counts the number of $1$s in a specific dimension.
We rather reuse the generic \MAN module that replaces the multibit hypervector components with binarized hypervector components by incrementally applying an approximate majority gate for every training example.  
Such a binarized back-to-back bundling enables the representational system to continuously stay in the binary space that is essential for efficient on-chip learning during the course of online learning. %
%

(3) The common denominator of all architectures of HD computing is the extensive use of distance computation in the associative memory that typically takes $\mathcal{O}(D)$ cycles per every event of classification.
We propose associative memories to significantly reduce the classification latency to single cycle.

(4) We perform a design space exploration of our library modules for an application which recognizes hand gestures from four EMG senors (Section~\ref{sec:design_space}).
It shows that functionally equivalent HD architectures can be composed achieving up to $2.39\times$ area saving, or $2337\times$ throughput improvement.
The Pareto optimal HD architecture is fully synthesized on only 18340 CLBs of the Xilinx\textsuperscript{\textregistered} UltraScale\textsuperscript{\texttrademark} FPGAs, and shows simultaneous $2.39\times$ area and $986\times$ throughput improvements compared to a baseline HD architecture.
%
%
%

\section{Background}
\label{sec:background}
HD computing is rooted in the observation that key aspects of human memory, perception and cognition can be explained by the mathematical properties of hyperdimensional spaces, and that a powerful system of computing can be built on the rich algebra of hypervectors~\cite{HD09}. 
A further motivation is the fact that brains compute with \emph{patterns of neural activity} that are not readily associated with numbers.  
In fact, recognizing the very size of the brain’s circuits, we can model neural activity patterns with points in a hyperdimensional space. 
Computing in hyperdimensional space is understood partly in terms of the linear algebra and probability of artificial neural nets, and partly in terms of the abstract algebra and geometry of hyperdimensional spaces. 
Groups, rings, and fields over hypervectors become the underlying computing structure, with permutations, mappings, and inverses as primitive computing operations, and with randomness as a way to label new objects and entities. 

Hypervectors are $D$-dimensional, \emph{holographic}, and (pseudo)random with i.i.d. components.
%
%
It means that the contained information in a hypervector is distributed equally over all $D$ components: neither a component nor a subset of them have a specific meaning, hence the information degrades in relation to the number of failing components \emph{irrespective of their position}.
The high dimensionality yields a huge number of different, nearly orthogonal hypervectors in such space~\cite{SDM}. 
They can be mathematically manipulated for solving cognitive tasks, e.g., Raven's progressive matrices~\cite{BSC+SDM}, analogical reasoning~\cite{Dollar-Mexico}, and practical learning and classification tasks~\cite{TCAS17,Aditya,Rahimi2016,Rasti2016,IEDM_2016,ISSCC18,Rasanen2015con,VoiceHD_ICRC17,EMG-HD,ISCAS18,DAC18,TNNLS18,BICT17,MONET17,Rasanen14,Rasanen15,Robot_LearnBehHierarchies,Robot_LearnByDem,PingChen_TCAS17}.
Examples of such computing include Holographic Reduced Representation~\cite{PlateBook,PlateTr}, Binary Spatter Code~\cite{BSC96}, Multiply-Add-Permute architecture~\cite{MAP98}, Random Indexing~\cite{Randomindexing}, and Semantic Pointer Architecture Unified Network~\cite{SPAUN14}, collectively referred to as Vector Symbolic Architecture~\cite{VSA03}. 
They differ in the type of components, and the types of operations, however, the key properties are shared by hypervectors of many kinds, all of which can serve as the computational infrastructure.
To ease the hardware realization, we focus on Binary Spatter Code (BSC), where the components of hypervectors are binary and dense, meaning the probability of having a $1$ or a $0$ is equal ($p=1/2$)~\cite{BSC96}.
%
%
%
\subsection{Measure of Similarity}
\label{sec:algorith_hamming}
\begin{figure}[b!]
	\centering
	\includegraphics[width=0.65\linewidth]{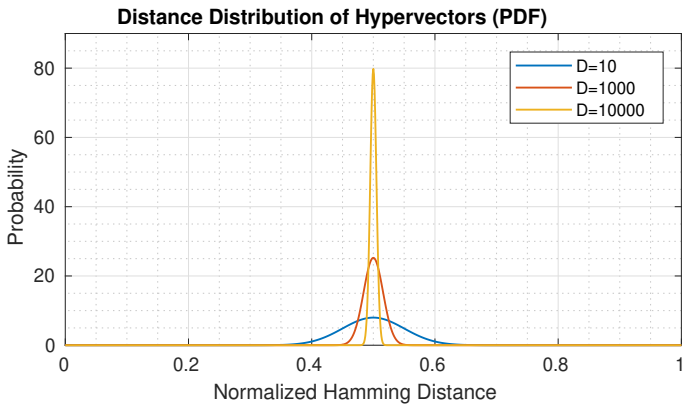}
	\caption{Normalized Hamming distance distribution of hypervectors in $D$-dimensional spaces. As the dimensionality increases, the standard deviation ($1/(2\sqrt{D})$) of the normalized distance distribution between two random hypervectors decreases. This implies that the probabilty of two random hypervectors lying about $d \approx 0.5$ apart from each other increases with the dimension $D$.}
	\label{fig:distances_in_hd}
\end{figure}
Using BSC, $\{0,1\}^D$, the similarity between two hypervectors is given by the number of components at which they differ, the so-called \textit{Hamming distance}.
We use the normalized version of this metric by dividing by $D$ denoted as: $d(X,Y): \{0,1\}^D \times \{0,1\}^D \rightarrow [0,1]$ to express the distance on a real scale of 0 to 1.
Figure~\ref{fig:distances_in_hd} shows the normalized Hamming distance distribution of hypervectors in $D$-dimensional spaces where $D \in \{100,1000,10000\}$. 
As we go to higher dimensions from $D=100$ to $D=10,000$, we observe an outstanding property: most points are $D/2$ bits apart from each other, which yields a normalized Hamming distance of $d \approx 0.5$, and stands for two nearly orthogonal hypervectors. 
%
%
This stems from the binomial distribution for $p=1/2$ and $n=D$, where $D/2$ is the mean. 
Correlated hypervectors yield $d \approx 0$ whereas $d \approx 1$ implies anti-correlation~\cite{HD09}.

\paragraph{Orthogonality Condition}
\label{sec:algorithm_orthogonality}
When approximating the discrete binomial distribution with the continuous normal distribution, its standard deviation is $\sqrt{D}/2$. 
According to the normal distribution, $\approx 68.2\%$ of the space lies within one standard deviation from the mean or within $\sqrt{D} \pm 1$ standard deviations from a point in the hyperdimensional space~\cite{SDM}.
If we increase the range to 6 standard deviations, already $\approx 99.9999998\%$ of the space lies within that range. 
This marks our orthogonality threshold as $d_{orthogonality} = \frac{\sqrt{D} \cdot (\sqrt{D} \pm 6)}{2 \cdot D}$
which states that with a chance of $\approx 99.9999998\%$ two random hypervectors exhibit a normalized Hamming distance in the aforementioned range.
For $D=10,000$ this yields a range between $0.47$ and $0.53$~\cite{SDM}.
In other words, almost all the space lies at approximately the mean distance of [0.47,0.53] from a chosen random point; this implies that for any significant deviation from distance 0.5, the distribution quickly becomes very sparse.

\subsection{HD Arithmetic Operations}
The HD algorithm starts by choosing a set of seed hypervectors as initial items.
They are stored in a so-called item memory (IM) as a symbol table or dictionary of all the
hypervectors defined in the system.
They stay fixed throughout the computation, and they serve as seeds from which further
representations are made.
HD computing builds upon a well-defined set of operations with the seed hypervectors~\cite{HD09}. 
These arithmetic operations are used for encoding and decoding patterns. 
The power and versatility of arithmetic derives from the fact that addition and multiplication form an algebraic field, and permutation of hypervector components takes it beyond both arithmetic and linear algebra. 
%

\paragraph{Addition (Bundling)}
\label{sec:algorithm_bundling}
The sum of binary hypervectors is defined as the componentwise majority function (also called the median operator) with ties broken at random. 
This means, when adding an even number of hypervectors, in case of disagreement for a component (equal number of $1$s and $0$s), the majority is randomly chosen. 
It is denoted as $A \oplus B$.
The sum of two hypervectors stores information from both hypervectors, due to the mathematical properties of vector addition, therefore the operation is also called \emph{bundling}. 
Bundling two hypervectors yields a hypervector which is similar to both of them, hence it is well-suited for representing sets or multisets.
However, when breaking ties at random, the bundling operation becomes non-causal. 
Furthermore, the bundling is commutative but not associative and is only approximately invertible.
\paragraph{Multiplication (Binding)}
The product of two binary hypervectors is defined as the componentwise XOR or ``addition modulo 2'', and is denoted as $A \otimes B$.
The resulting hypervector is dissimilar (orthogonal) to both its constituent hypervectors, which is why multiplication is well-suited for \emph{binding} two hypervectors.
Binding is commutative, associative and distributes over bundling. 
The operation can be inverted and also preserves distances between hypervectors, meaning two similar hypervectors (after binding) are mapped to equally similar ones.
\paragraph{Permutation}
\label{sec:algortithm_perm}
The third operation, denoted $\rho(A)$, is the permutation operation, which shuffles a hypervector's components by rotating it in space. 
It is implemented as a cyclic shift by one position.
Permuting a hypervector produces a dissimilar, pseudo-orthogonal hypervector, which can be exploited to bypass the commutativity of the other operations. 
This is crucial when storing sequences, where e.g., a-b-c should be distinguishable from b-c-a.
Permutation is invertible and preserves distances. 
It distributes over both bundling and binding.

These three operations can be combined to encode structures such as variable/value records, sequences, and lists---essentially any data structure. 
For example, let us consider three variables $x$, $y$, $z$ and their values $a$, $b$, $c$. 
Each of them is mapped to a (random) hypervector $X$, $Y$, $A$, $B$ etc., which are stored in the IM.
Then, the entire of a record is encoded to a single hypervector by binding each value to its variable and bundle them to form the holistic record:
$R = (X \otimes A) \oplus (Y \otimes B) \oplus (Z \otimes C)$.
To find the value of $x$, we unbind the record with the inverse of $X$ (which is $X$ itself), $\tilde{A} = X \otimes R$ which gives us a hypervector $\tilde{A}$ as noisy version of $A$. 
After comparing it with the hypervectors that are stored in the AM, we find $A$ to be the most similar one (i.e., the lowest Hamming distance), and thus the sought value.


\section{Learning and Classifying Multichannel Biosignals with HD Computing}
\label{sec:biosignals_with_HD}
In this section, we describe how to use HD computing for learning and classification tasks.
%
%
We focus on wearable biosignal processing applications with multichannel noisy sensors for which HD computing achieves faster training and lower energy consumption and memory than SVMs~\cite{DAC18,BioCAS18}.
One application example includes recognizing hand gestures from a stream of EMG sensors to control a prosthetic device~\cite{EMG-HD,DAC18}. 
The performance of HD computing however depends on good design of a network architecture that demands a reconfigurable (FPGA) fabric to efficiently arrange the HD primitive operations based on the given task.  
We present a generic architecture to project multichannel sensory inputs from original representation to hyperdimensional space, where the arithmetic operations are combined to learn and classify examples.
While this paper focuses on EMG signals, other streaming multichannel sensor data such as ECoG~\cite{BioCAS18},
EEG~\cite{BICT17,MONET17},
ExG~\cite{Proc18},
speech~\cite{Rasanen2015con,VoiceHD_ICRC17}, smell~\cite{PingChen_TCAS17} can be equally applicable.

The dataset~\cite{EMG-HD} used in this paper is based on a four-channel EMG data acquisition, among five subjects, for the most common hand gestures in daily life.
The selected gestures are: closed hand, open hand, 2-finger pinch, point index, and the rest position.
The recording is composed of 10 trials of every gestures three seconds each.
We use 25\% of this dataset for training that can be performed online.
The gestures are sampled at 500\,Hz, followed by a low pass filter, and an envelope signal extraction; ~\cite{EMG-HD} provides further details about the setup. 

\subsection{HD Architecture}
\begin{figure}[t]
	\centering
	\includegraphics[width=\linewidth]{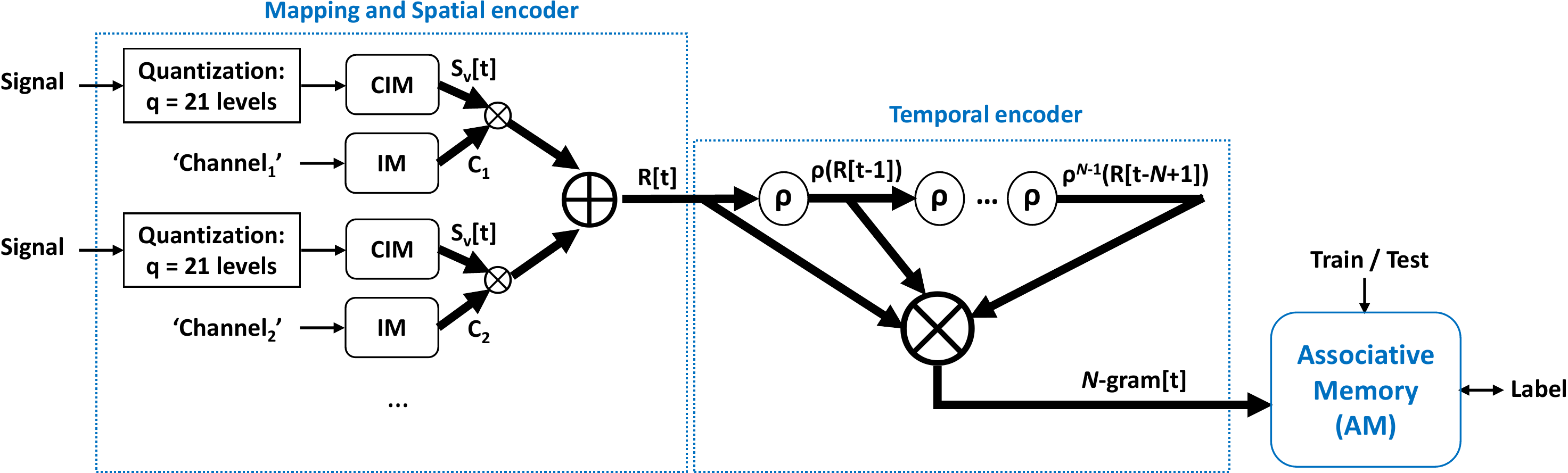}
	\caption{Example of an HD architecture for hand gestures learning and classification from EMG biosignals.}
	\label{fig:encoding_scheme}
\end{figure}
As shown in Figure~\ref{fig:encoding_scheme}, an HD architecture consists of three main modules: mapping and spatial encoder, temporal encoder, and associative memory. 
The mapping and encoding modules intend to capture information that can be extracted from the inputs (i.e., the enveloped EMG signals), into a hypervector representing a gesture. 
Gesture hypervectors, extracted from various trials, are bundled to form a \emph{prototype} hypervector representing a \emph{class} of gestures. 
The associative memory (AM) stores a prototype hypervector for every class, which contains the encoded information of all labelled inputs during the training phase.
During inference, classifying input data is carried out by comparing the unlabelled encoded hypervectors with all stored prototype hypervectors, and returning the label of the most similar one.
\subsection{Mapping and Spatial Encoder}
\label{sec:algorithm_cim}
\begin{figure}[b]
	\centering
	\begin{subfigure}{0.45\textwidth}
		\centering
		\includegraphics[scale=0.6]{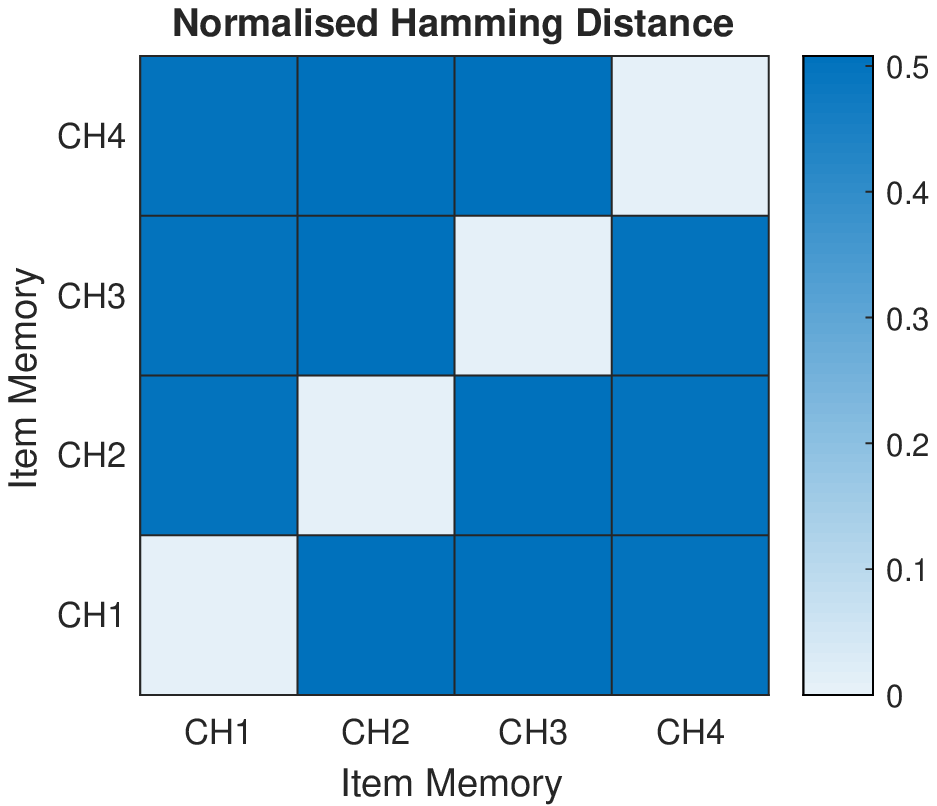}
		\caption{The IM with four channels}
		\label{fig:im_heatmap}
	\end{subfigure}
	\hspace{0.25cm}
	\begin{subfigure}{0.45\textwidth}
		\centering
		\includegraphics[scale=0.5]{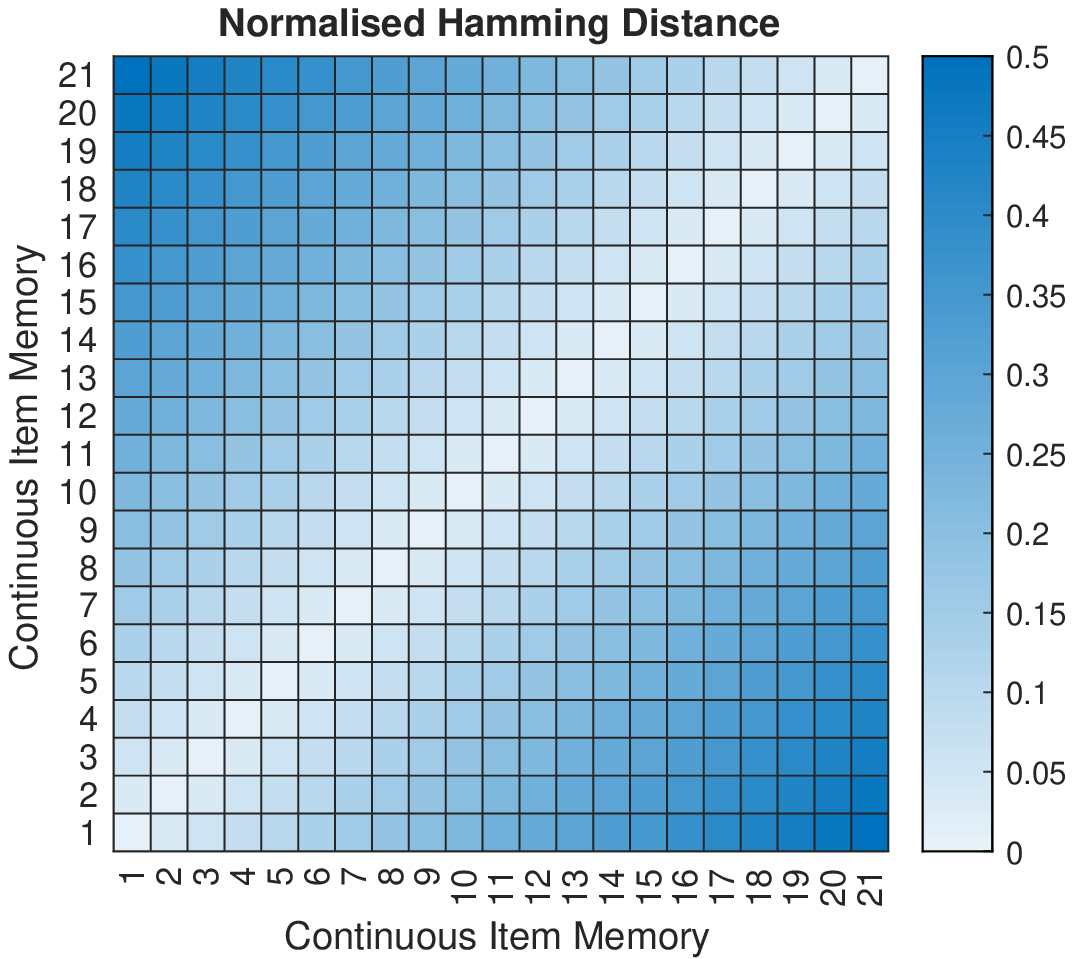}
		\caption{The CIM with $q=21$ levels}
		\label{fig:cim_heatmap}
	\end{subfigure}
	\caption{Similarity, depicted as a heatmap, between hypervectors of the IM (a) and the CIM (b).}
\end{figure}

First, the analog EMG signals have to be quantized to $q$ discrete levels, where $q$ indicates the resolution of the signal. 
In analogy to the record example in the previous section, the different EMG channels represent the variables or fields, and the discretized signals represent the values of the variables. 
All channels are treated as separate and independent, therefore we allocate each one a random and thus orthogonal hypervector, which are fixed throughout the computation in the IM: $C_1 \perp C_2 \perp C_3 ... \perp C_n$.
Figure~\ref{fig:im_heatmap} shows the IM with four channels.

Each of the channel variables has a corresponding value, i.e., the discretized signals. 
When mapping quantities from the discrete number space to the hypervector space, we want to retain their similarity: e.g., with a resolution of $q=21$ levels, a value of 5 is only slightly larger than a value of 4, hence their allocated hypervectors shall not be orthogonal~\cite{EMG-HD}. 
For mapping such quantized or even continuous values into hypervectors various techniques can be used including thermometer codes, locality-sensitive hashing, or generally, random projection~\cite{RandomProjection_review}.
We use the following simple method to map the values to a continuous vector space.
A random seed hypervector is taken for the smallest value and the hypervectors for the other levels are generated such that they are gradually further away from the seed up to the largest value, whose hypervector is orthogonal to the seed. 
We can accomplish this by randomly choosing $D/2$ components of the seed and split them into $q-1$ groups which equally contain $(D/2)/(q-1)$ components. 
The hypervectors are then generated from the seed by taking one group after the other and flipping their components. 
For the last hypervector, exactly $D/2$ components are flipped, making it orthogonal to the seed.
These generated \emph{signal} hypervectors are denoted by $S_v$ where $v \in [0,q-1]$, that are stored in the so-called \textit{continuous item memory} (CIM).
Figure~\ref{fig:cim_heatmap} illustrates a CIM with $q=21$: 
$
d(S_n,S_{n+i}) = 0.5 \cdot \frac{i}{q-1} 
$
hence
$
d(S_0,S_{q-1}) = 0.5 
$
.


As mentioned in Section \ref{sec:algortithm_perm}, we aim to bind the values to their variables and bundle them to form a holistic record ($R$) to capture spatial information between all channels.
The signal hypervector of a channel at time $t$, is denoted by $S_v[t]$ where $v \in [0,q-1]$.
Hence, a record is computed for a given time-aligned sample of all channels:
$R[t] = (C_1 \otimes S_v[t]) \oplus (C_2 \otimes S_v[t]) \oplus (C_3 \otimes S_v[t]) \oplus (C_4 \otimes S_v[t])$.
As shown in Figure~\ref{fig:encoding_scheme}, this record contains the signal information of all channels, while distinguishing the source of the signals (i.e., the channels).

\subsection{Temporal Encoder}
\label{sec:algorithm_temporal_encoding}
We can encode sequences by using the permutation operation $\rho$. 
Hence, we can capture not only the spatial correlation across the channels, but also the temporal correlation between subsequent samples. 
We call a sequence of $N$ record hypervectors as an $N$-gram hypervector.

As already mentioned, a sequence of hypervectors can be encoded uniquely by permuting the hypervectors before binding them. 
The sequence is encoded by rotating the first spatial record $N-1$ times, the second $N-2$ times, and the $(N-1)$th only once.
The $N$th hypervector is untouched (not permuted).
These new hypervectors are finally bound to an $N$-gram (see Figure \ref{fig:encoding_scheme}).
%
%
For large $N$-grams, this becomes: $N\textrm{-gram}[t] = \prod_{i=0}^{N-1}\rho^i(R[t-i])$.
An $N$-gram contains the spatial information of $N$ subsequent samples with different timestamps, making it a \textit{spatiotemporal} hypervector.

\subsection{Learning and Classification in Associative Memory}
In a typical training setting, a set of labelled examples is provided per every class.
By encoding the sensory data, a current gesture example is represented by an $N$-gram[$t$] hypervector.
The HD architecture learns from these $N$-gram hypervectors that are produced over time.
A number of $N$-gram hypervector examples (e.g., $k$) with the same label are bundled to produce a prototype hypervector representing the class of interest: $P_{\textrm{Label}_i} = N\textrm{-gram}_{\textrm{Label}_i}[t]\oplus 
...\oplus
N\textrm{-gram}_{\textrm{Label}_i}[t+k]$.
Once training is done, the binarized prototype hypervectors are stored in the AM as \emph{learned patterns}.
This \emph{temporal} bundling of $N$-grams over the course of training requires $D$ counters and thresholders to implement the majority function.
%

As soon as the AM is trained for each class, it can identify the corresponding class of an unlabelled $N$-gram, which is called a \textit{query hypervector}. 
More specifically, the AM computes the Hamming distance between the query hypervector and each of its prototype hypervector.
It then selects the highest similarity and returns its associated label.
As shown in Figure~\ref{fig:encoding_scheme}, the same construct is reused during inference, the only difference is that during training the prototypes are written into the AM while during inference they are read and compared with the query.



\section{Hardware Optimizations of Dense Binary HD Computing}
\label{chap:optimisations}
\begin{table}[b!]
	\centering
	\caption{Cycles per data item (CPDI) orders of the different library modules.}
	\label{tab:cpdi}
	\begin{subtable}[t]{0.31\textwidth}
		\centering
		\caption{Spatial encoder modules.}
		\label{tab:cpdi_spatial_encoder}
		\begin{tabular}{ll}
			\toprule
			\textbf{Module}	& \textbf{CPDI} \\
			\hline
			LUT 			& $\mathcal{O}(1)$ \\
			CA 				& $\mathcal{O}(n_\textrm{channels})$ \\
			MAN				& $\mathcal{O}(n_\textrm{channels})$ \\
			\bottomrule
		\end{tabular}
	\end{subtable}
	\hspace{0.25cm}
	\begin{subtable}[t]{0.31\textwidth}
		\centering
		\caption{Temporal encoder modules.}
		\label{tab:cpdi_temporal_encoder}
		\begin{tabular}{ll}
			\toprule
			\textbf{Module}	& \textbf{CPDI} \\
			\hline
			BC 				& $\mathcal{O}(1)$\footnote{This holds only for inference. During training, the order is of the number of training samples.} \\
			B2B				& $\mathcal{O}(1)$ \\
			\bottomrule
		\end{tabular}
	\end{subtable}
	\hspace{0.25cm}
	\begin{subtable}[t]{0.31\textwidth}
		\centering
		\caption{Associative memory modules.}
		\label{tab:cpdi_associative_memory}
		\begin{tabular}{ll}
			\toprule
			\textbf{Module}	& \textbf{CPDI} \\
			\hline
			BS 				& $\mathcal{O}(D)$ \\
			CMB 			& $\mathcal{O}(1)$ \\
			VS				& $\mathcal{O}(n_\textrm{classes})$ \\
			\bottomrule
		\end{tabular}
	\end{subtable}
\end{table}
In this section, we present the main contributions of the paper.
We present our techniques to optimize hardware realization of HD computing suitable for CMOS fabrics.
HD computing demands a large amount of bits to be stored for each data item that further poses a memory bandwidth issue, for instance the IP RAMs of FPGAs are optimized for usually no more than 72 bits in parallel~\cite{ds_xilinx_us}. %
Storing or loading one hypervector in this fashion would require hundreds of cycles.
Accordingly, optimizing the architecture of HD computing should focus on minimizing the number of stored hypervectors.
Furthermore, the bitwise operations need to be kept as simple as possible, since they are replicated over the whole dimension of a hypervector. 
Most architectural constructs are shared among various HD classifiers and thus the optimizations virtually concern all HD computing applications.

As a result of various hardware optimizations, we introduce a synthesizable VHDL library of fully configurable modules which comprises different implementations.
The VHDL library consists of interchangeable modules including three types of spatial encoder, two types of temporal encoder, and three types of AM, that are listed in Table~\ref{tab:cpdi}.
A functioning HD architecture can be configured by connecting one type of each of the modules in series.
The modules operate independently and pass hypervectors after synchronizing via handshake signals.
They all differ greatly in area and throughput, where the number of cycles needed to process a data item (CPDI) has the biggest influence on throughput.
Table~\ref{tab:cpdi} shows the CPDI for the different modules.


\subsection{Mapping Multichannel Sensory Inputs}
\label{sec:opt_mapping}
Mapping the input data of more than one channel to the hyperdimensional space can be done in a parallel fashion as shown in Figure~\ref{fig:encoding_scheme}. 
The required memory for the IM and the CIM is $n_c \times q \times D$ where $n_c$ is the number of channels, $q$ is the quantization of input signal, and $D$ the hypervector dimension. 
For the EMG task (see Figure~\ref{fig:encoding_scheme}) this would be equal to $840$ kbits to only store the seed hypervectors. This poses limitations when a large number of channels~\cite{ISCAS18} or input quantization is used.
A first step is to trade the high throughput against a smaller memory footprint by sharing the resources. 
\subsubsection{Rematerialization: Replacing CIM with \MAN}
\label{sec:rep_luts}
A single CIM implemented as a lookup table requires $q \times D$ bits of storage.
To reduce this memory footprint we can exploit the holographic nature of HD representation: the individual bits in a hypervector do not represent anything. 
What is important is the relation or similarity between two hypervectors.
A hypervector can be altered or ``manipulated'' to a different hypervector by switching certain bits as a function of the similarity that we want to establish. 
For example, to obtain an orthogonal hypervector, we have to switch half of its bits (which ones does not matter), whereas to obtain a similar hypervector, we only switch a (small) portion of the bits (see Section~\ref{sec:algorith_hamming}).

\begin{figure}[t]
	\centering
	\includegraphics[scale=0.65]{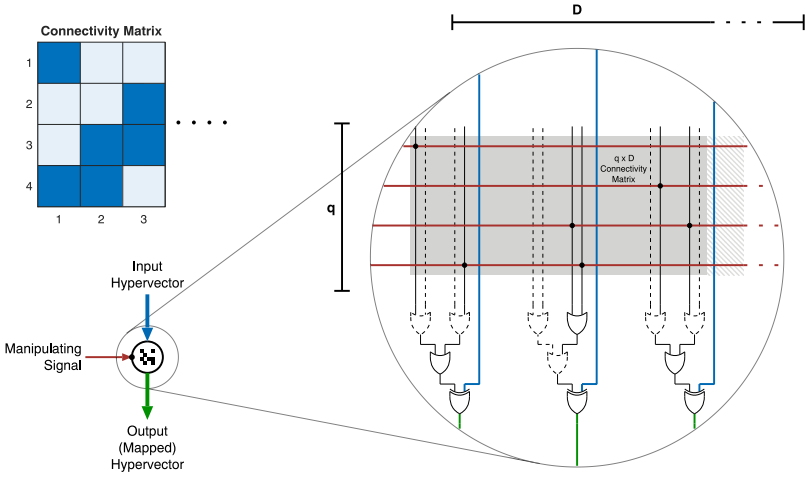}
	\caption{The hypervector manipulator (\MAN) module and its symbol representation. The connectivity matrix serves as an example. The other symbols used throughout this paper can be found in Figure~\ref{fig:symbols}.}
	\label{fig:manipulator}
\end{figure}

\begin{figure}[b]
	\centering
	\includegraphics[width=\linewidth]{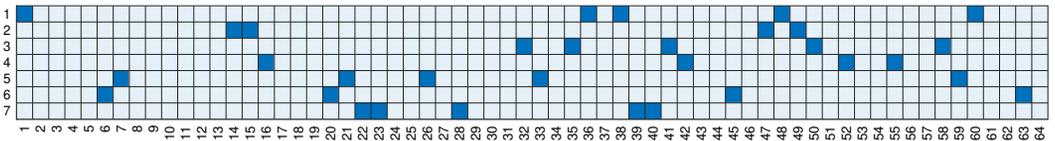}
	\caption{Example of a connectivity matrix to map an input with $q = 8$ to a hypervector of dimension $D = 64$.}
	\label{fig:ex_cim_con_mat}
\end{figure}

Manipulating hypervectors in a controlled manner can replace complex constructs throughout the whole architecture. 
For this purpose, a generic hypervector manipulator (\MAN) module is designed (Figure~\ref{fig:manipulator}), which can be configured in depth and width, and is fixed by a \textit{connectivity matrix}, which determines the connections between wires. 
An example connectivity matrix used for mapping is shown in Figure \ref{fig:ex_cim_con_mat}.

\begin{figure}[t]
	\centering
	\includegraphics[scale=0.7]{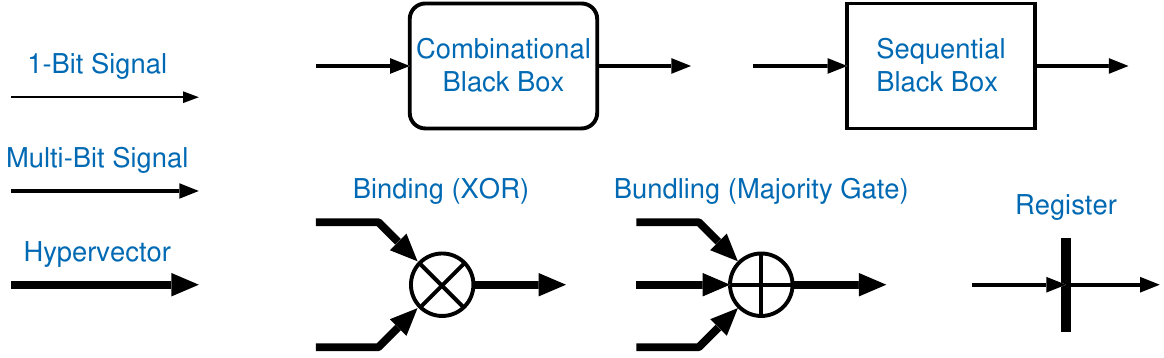}
	\caption{The symbols used for schematic drawings throughout this paper.}
	\label{fig:symbols}
\end{figure}

Every cell of the connectivity matrix affects, whether a certain bit of the input hypervector can be switched by a bit (or even several bits) of the \textit{input manipulator}. 
The \MAN module is a simple combination of OR and XOR gates.
If a cell $(m,n)$ of the connectivity matrix is set to $1$, the $m$-th bit of the input manipulator can affect the $n$-th bit of the input hypervector: when the $m$-th bit of the input manipulator is logically high it toggles the $n$-th bit of the input hypervector. 
The number of $1$s in a row of connectivity matrix also represents how dissimilar the output hypervector will be to the input hypervector when the input manipulator bit of that row is logical high: the fewer the number, the more similar.

\begin{figure}[b]
	\centering
	\includegraphics[scale=0.8]{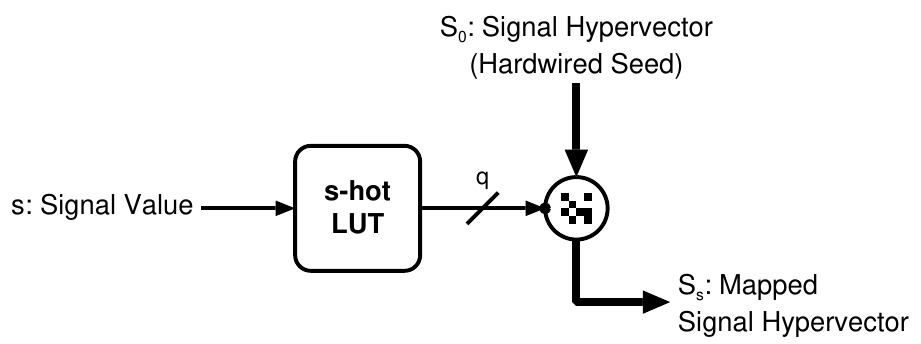}
	\caption{Data dependency graph of the \MAN module to replace a CIM.}
	\label{ddg:mapping}
\end{figure}

As described in Section~\ref{sec:algorithm_cim}, ``close'' input values are mapped to similar hypervectors using a CIM.
This CIM can be replaced by a \MAN module that produces similar hypervectors according to the input value.
First, the quantized input value in binary representation is mapped to an $s$\textit{-hot} representation (by e.g., a small lookup table), where $s$ is the input/signal value (see Figure~\ref{ddg:mapping}). 
This $s$-hot code serves as the input manipulator, and gradually switches more and more bits of a \emph{seed} input hypervector as the input value goes higher, and eventually produces an orthogonal hypervector when all $q$ bits are hot ($q$ is the quantization).
This allows to rematerialize desired hypervectors from a seed by keeping track of the input value.

Which bits are switched is chosen randomly (without the possibility to choose a bit twice), only the number of bits per ``input quantum''---represented by a row in the connectivity matrix---is determined. 
It is equal to $D / 2 / (q-1)$.
Moreover, every input hypervector bit can only be switched by one input manipulator bit. 
This results in a \MAN module containing only XOR gates. 
The input hypervector that is manipulated is a constant seed hypervector ($S_0$) which represents the lowest input value, or 0-hot. 
This seed hypervector is simply hardwired connections to source and ground. 
Summing up, the whole continuous item memory, or CIM, is replaced with a rather small $s$-hot lookup table memory of size $q \times q$, some wires, and $D/2$ XOR gates.


%
\subsubsection{Reproducing IM with Cellular Automata}
\label{sec:opt_ca}
As mentioned in Section~\ref{sec:algorithm_cim}, we account for the spatial multichannel information to determine which channel the data originated from. 
This is done by binding a channel hypervector, that is unique for every channel, with the signal hypervector. 
The channel hypervectors are typically stored in the IM with a memory of size $n_c \times D.$
When mapping the input data in the parallel fashion, the IM can be replaced by hard wires tied to source and ground since the channel hypervectors are constant. 
However, with the serial mapping, they need to be stored in the IM.

\begin{figure}[b]
	\centering
	\includegraphics[width=0.6\textwidth]{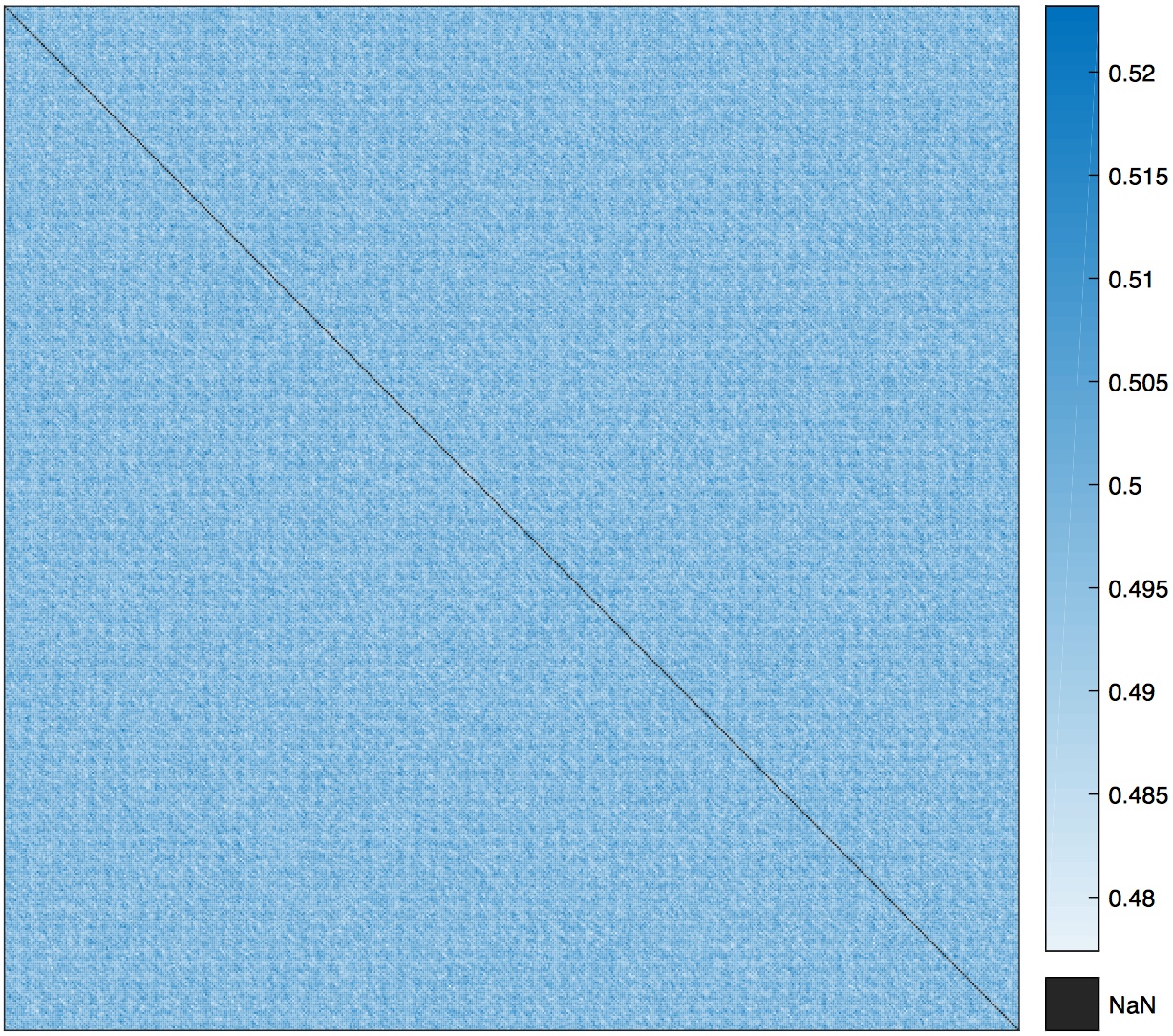}
	\caption{A heat map showing the orthogonality (normalized Hamming distance) between hypervectors produced by the CA (rule 30) over 500 cycles. Each dot $(x,y)$ on the graph shows the Hamming distance between the hypervector produced in cycle $x$ and the one produced in cycle $y$. As shown in the minimum and maximum values of the color scale on the right, the orthogonality condition from Section~\ref{sec:algorithm_orthogonality} is met.}
	\label{fig:heatmap_ca}
\end{figure}

\begin{figure}[t]
	\centering
	\includegraphics[scale=0.8]{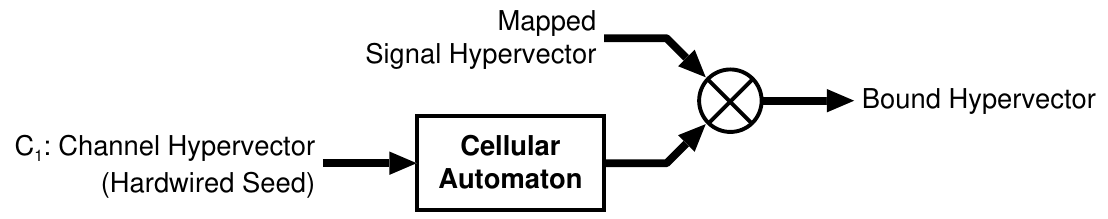}
	\caption{Data dependency graph of the spatial encoding architecture with the cellular automaton (CA).}
	\label{ddg:cellular_automaton}
\end{figure}

One way to replace the IM is by using a one-dimensional cellular automaton (CA) with a \textit{neighborhood} of 3, applying \textit{rule 30}~\cite{ca30}. 
This rule exhibits chaotic behaviour that is well-matched to produce a sequence of (quasi-)random hypervectors. 
When using a CA with $D$ cells and a random hypervector as initial state, it generates (quasi-)random and orthogonal hypervectors every cycle (see Figure~\ref{fig:heatmap_ca}).
By resetting the CA registers, the same sequence can be reproduced (i.e., rematerialized) over and over.
This allows us to replace the IM (see Figure \ref{ddg:cellular_automaton}) by only defining the initial state of the CA as a seed hypervector and letting it generate the other orthogonal hypervectors\footnote{In case the ``randomness'' of rule 30 is not enough, the neighbourhood can be extended to form a more complex CA as in \cite{ca_5neighbours}.} for the rest of the channels.
%
Thanks to the chaotic behaviour of the CA, this approach works for virtually any number of channels: clocking the CA for 500 cycles produces the channel hypervectors for 500 channels only from the initial state hypervector (see Figure~\ref{fig:heatmap_ca}).


Although the gate logic required for each cell in CA is quite simple---only consisting of 3 inverters, 4 two-input AND gates and 2 two-input OR gates---it is still replicated $D$ times. 
When looking for a solution to generate orthogonal hypervectors at relatively low costs, CA are an excellent choice, whereas when looking for an optimal solution for spatial encoding, further improvement can be done as described in the following section.
\subsubsection{Replacing Both IM and CIM with \MAN}
\label{sec:opt_man}
\begin{figure}[b]
	\centering
	\includegraphics[width=\linewidth]{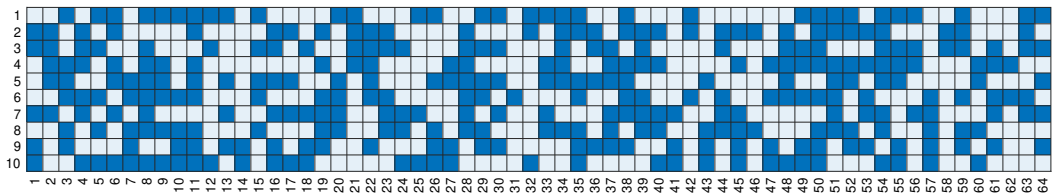}
	\caption{Example of a connectivity matrix to replace the IM for a hypervector dimension $D=64$ bits and 10 input channels.}
	\label{fig:ex_im_con_mat}
\end{figure}

%
%
The \MAN module in Section~\ref{sec:rep_luts} can also be applied to replace the IM. 
Instead of storing the channel hypervectors, their patterns can be incorporated in the connections of the \MAN module. 
The connectivity matrix in this case is identical to an IM and has an average of $D/2$ $1$s per row as shown in Figure~\ref{fig:ex_im_con_mat}. 
Feeding signal hypervector to the second \MAN module and setting one bit of its input manipulator logical high at a time yields the same outcome as binding the signal hypervector with a channel hypervector (see Figure~\ref{fig:spatial_encoder_man}).

The second \MAN module (replacing the IM) requires more gates due to its dense connections than the first one (replacing the CIM).
The chance that a channel hypervector switches a certain bit is $0.5$ (the probability of having a $1$ in a component), hence this yields an average of $n_c/2$ connections per column in the connectivity matrix (see Figure~\ref{fig:ex_im_con_mat}) which have to be OR-ed before going into the XOR gate. 
This operator per hypervector bit is replicated $D$ times to replace the whole IM.

\subsection{Spatial Encoding}
\label{sec:spatial_encoding}
The hypervectors that contain information of the input signal values and the channels should be bundled in the spatial encoder.
In Section~\ref{sec:algorithm_bundling}, the bundling operation is characterized as a method to store the information of multiple hypervectors in a single hypervector, called a record, which is similar to all of the input hypervectors. 
The information of a hypervector is contained in another as long as they do not violate the similarity condition (Section~ \ref{sec:algorithm_orthogonality}). 
Here, we investigate how well this task is accomplished by the majority function, and how it can be implemented in hardware and whether there are other approaches to achieve the same goal.

\subsubsection{The Three Problems of the Majority Function}
\begin{figure}[t]
	\centering
	\includegraphics[scale=0.8]{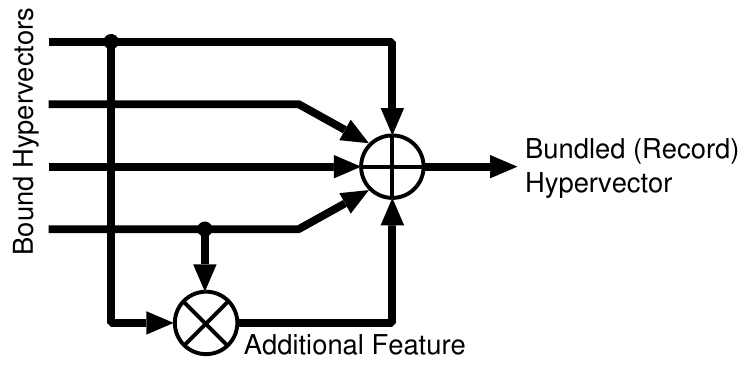}
	\caption{Data dependency graph of the bundling with even inputs using an additional feature.}
	\label{ddg:additional_feature}
\end{figure}

\paragraph{The Majority Function of an Even Number of Inputs}
\label{sec:maj_even_inp}
The majority function (or vote) for binary inputs is self-explanatory and only yields a clear result with an odd number of inputs. 
This is why the concept of \textit{braking ties at random} is introduced~\cite{HD09}, which makes the operation non-causal for an even number of inputs and is identical to bundling an additional random (and thus orthogonal) hypervector into the record. 
Therefore, two records, that are supposed to be equal, become (slightly) dissimilar. 
Instead of ``wasting'' said similarity, an additional hypervector can be introduced, that contains useful information, to break the ties.
In the case of bundling hypervectors from multichannel, useful information could come from an additional channel. 
If this is not an option, we can synthetically create that information. 
It should be ``useful'' in the sense, that it is unique for the given input and also causal. 
Binding a constant hypervector would lead to all output hypervectors being slightly similar to each other even if they are supposed to be orthogonal.
Instead, by simply binding any two of the input hypervectors (see Figure \ref{ddg:additional_feature}), we can create an \textit{additional feature}, which represents the input data and is useful as stated before. 

\begin{figure}[b]
	\centering
	\begin{subfigure}{0.45\textwidth}
		\centering
		\includegraphics[scale=0.75]{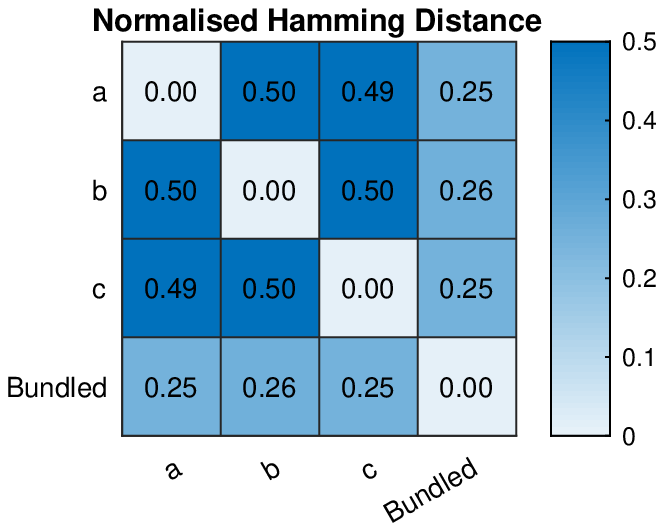}
		\caption{$a \perp b \perp c$}
		\label{fig:htmp_bun_3_gol_a}
	\end{subfigure}
	\hspace{0.25cm}
	\begin{subfigure}{0.45\textwidth}
		\centering
		\includegraphics[scale=0.75]{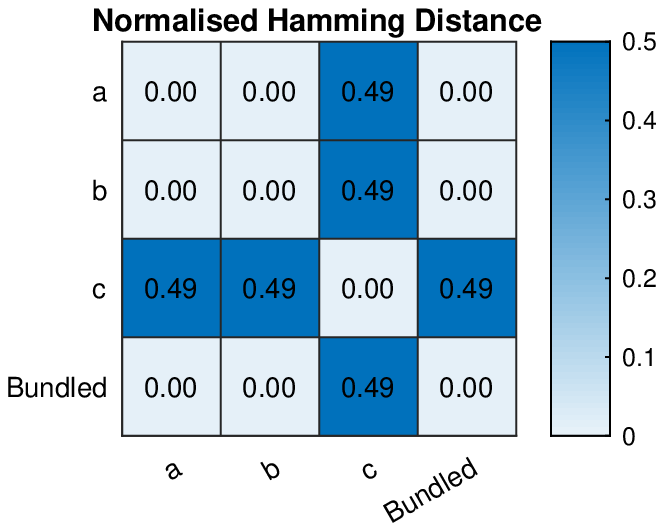}
		\caption{$a = b \perp c$}
		\label{fig:htmp_bun_3_gol_b}
	\end{subfigure}
	\caption{Similarity between three hypervectors and the bundled hypervector $a \oplus b \oplus c$ using majority vote.}
	\label{fig:htmp_bun_3_gol}
\end{figure}

\paragraph{Unfairness of the Majority Function}
Bundling hypervectors with the majority vote does not yield their mean hypervector but strongly tends to the majority of the hypervectors. 
This means, if we want to store the information of e.g., three hypervectors, where two of them are equal and the other is orthogonal, the information of the latter is lost entirely (see Figure~\ref{fig:htmp_bun_3_gol_b}).
The same situation occurs when bundling two sets of hypervectors to one record, where the sets are dissimilar to each other, but similar within.
The smaller set will not be recalled at all.
In Section~\ref{sec:approx_bundling}, another bundling approach will be presented, which is completely fair in this case.

When bundling only orthogonal hypervectors, this problem does not occur and the majority function is fair (Figure \ref{fig:htmp_bun_3_gol_a}). 
This rises the question of the ``capacity'' of the bundling operations (see Section~\ref{sec:capacity}).

\paragraph{Lack of Associativity}
When attempting to implement the bundling operation, one quickly comes across a mathematical property that is necessary to conduct an operation in an iterative manner: \textit{associativity}. 
The majority function lacks this property, meaning a set of hypervectors can only be bundled altogether, but not step by step: $a \oplus b \oplus c \neq (a \oplus b) \oplus c$.
Fortunately, one is not tied to mathematical properties, when it comes to the algorithmic and architectural implementation of an operation. The workaround lies in storing the current vote over an iteration.


\subsubsection{Bidirectional Saturating Counters as a Hardware Implementation of the Majority Function}
\label{sec:saturation_counter}
A naive approach to store the current majority vote would be to count the vote for $1$s and $0$s with two separate counters and compare their values to get the majority. 
This would require a memory of $2 \cdot D \times \ceil*{\log_2(n_c+1)}$ which, for only $n_c=4$ input channels in our EMG task would already yield
$60,000$ bits.

The two counters can be combined to a single one that counts up or down depending on the value of the current bit to reduce the memory to $D \times (\ceil*{\log_2(n_c+1)} + 1).$
The next big improvement is made by exploiting the random nature of orthogonal hypervectors. 
Observing a single component of the input vectors, the probability of a long sequence of either $1$s or $0$s is small, implying the counter usually does not have to count all the way up to the maximum possible vote, but stays within a certain range. 
Taking a counter with a fixed width and forcing it to saturate whenever it would traverse that range, assures that the vote is not passed to the other extreme, which occurs when letting it wrap around.

With this approach, the maximum accuracy of the majority function can be reached with a certain width of the counter.
For a hypervector dimension $D=10,000$ the maximum width is $5$ bits resulting in a memory of $50,000$ bits which is independent from the number of hypervectors to be bundled, and is maximally memory-saving for a large number of input channels.
The downside is the complexity of a saturating counter.
Due to the orthogonality of the hypervectors for bundling inside the spatial encoder, the saturating counter method is the preferred approach because of its large capacity and moderate complexity.


\subsection{Library: Spatial Encoder Modules}
\begin{figure}[h!]
	\centering
	\begin{subfigure}{\textwidth}
		\centering
		\includegraphics[scale=0.7]{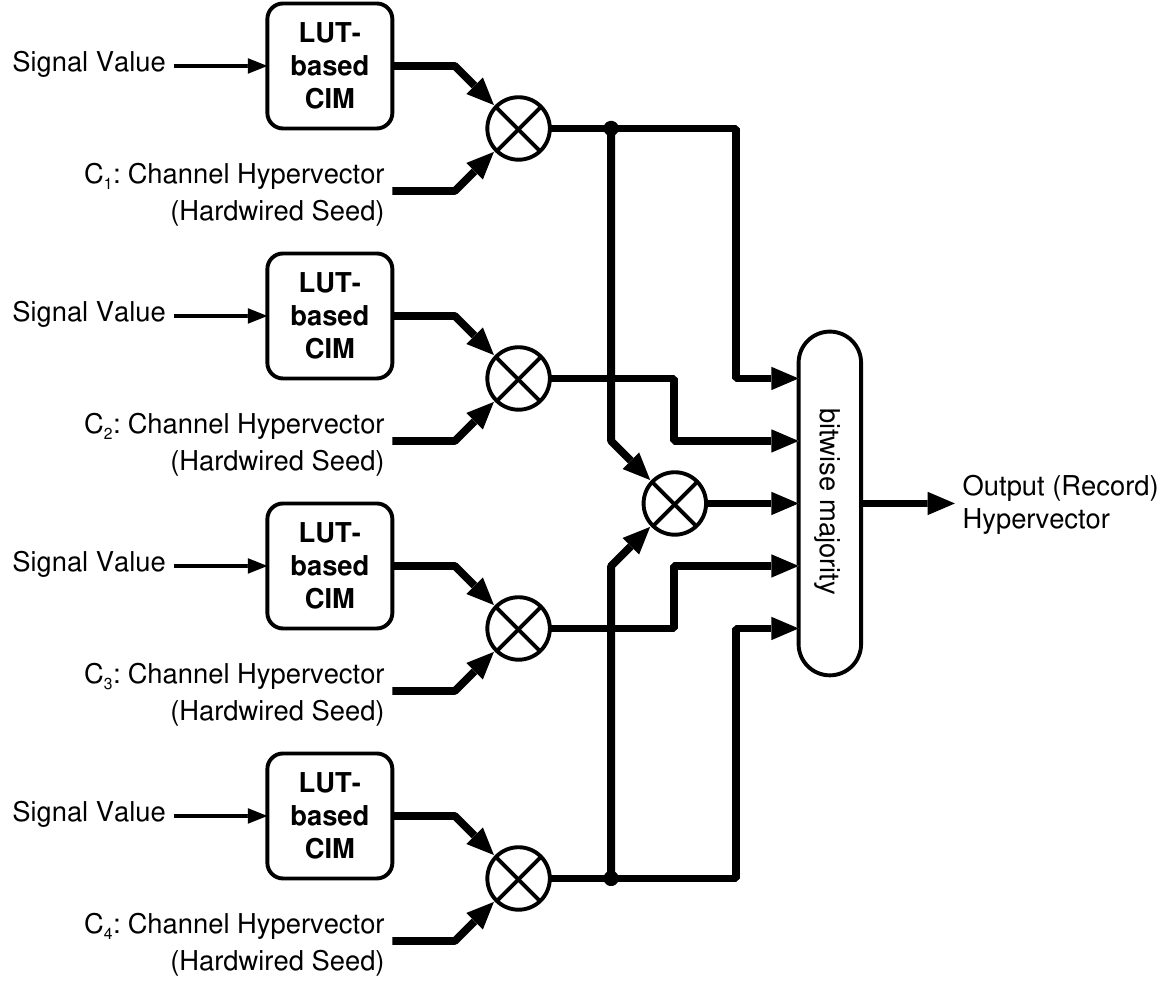}
		\caption{Lookup table (LUT): Parallel encoder with LUT-based CIMs and no IMs.}
		\label{fig:spatial_encoder_lut}
	\end{subfigure}
	\vspace{0.5cm}
	
	\begin{subfigure}{\textwidth}
		\centering
		\includegraphics[scale=0.7]{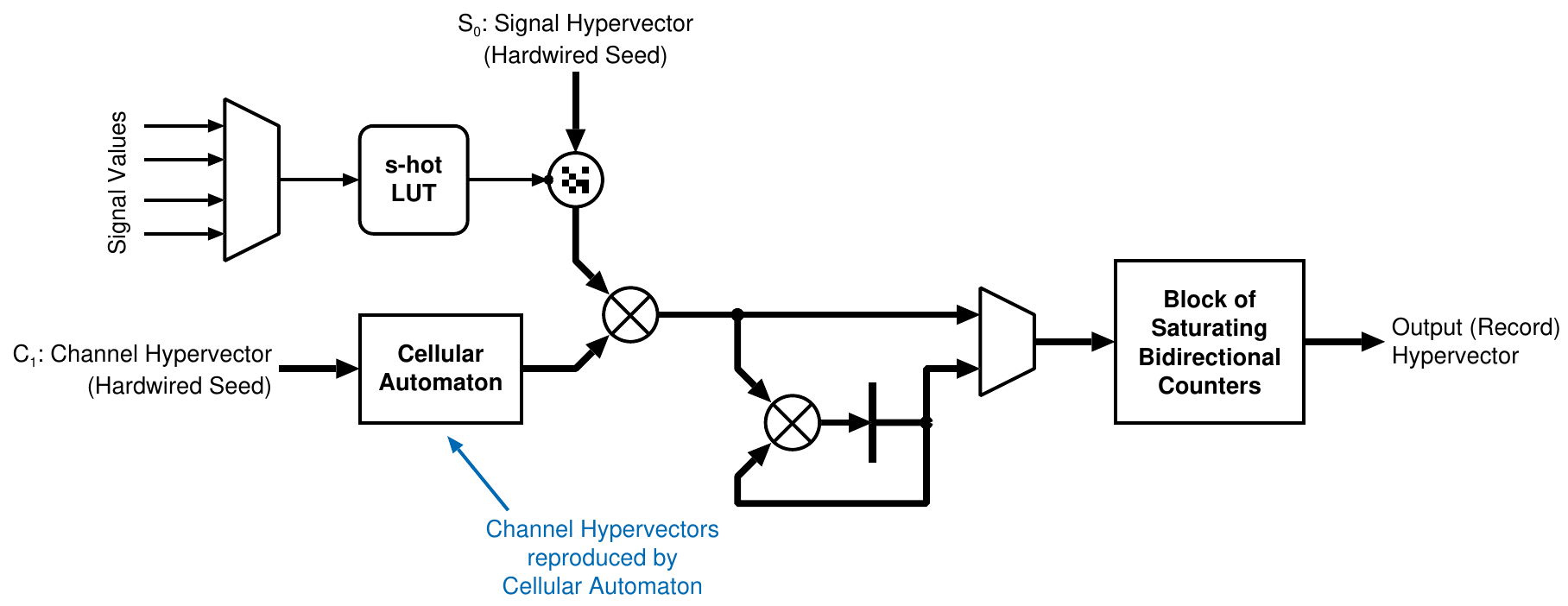}
		\caption{Cellular automaton (CA): Sequential encoder with CA (replacing IM), and \MAN module (replacing CIM).}
		\label{fig:spatial_encoder_ca}
	\end{subfigure}
	\vspace{0.5cm}
	
	\begin{subfigure}{\textwidth}
		\centering
		\includegraphics[scale=0.7]{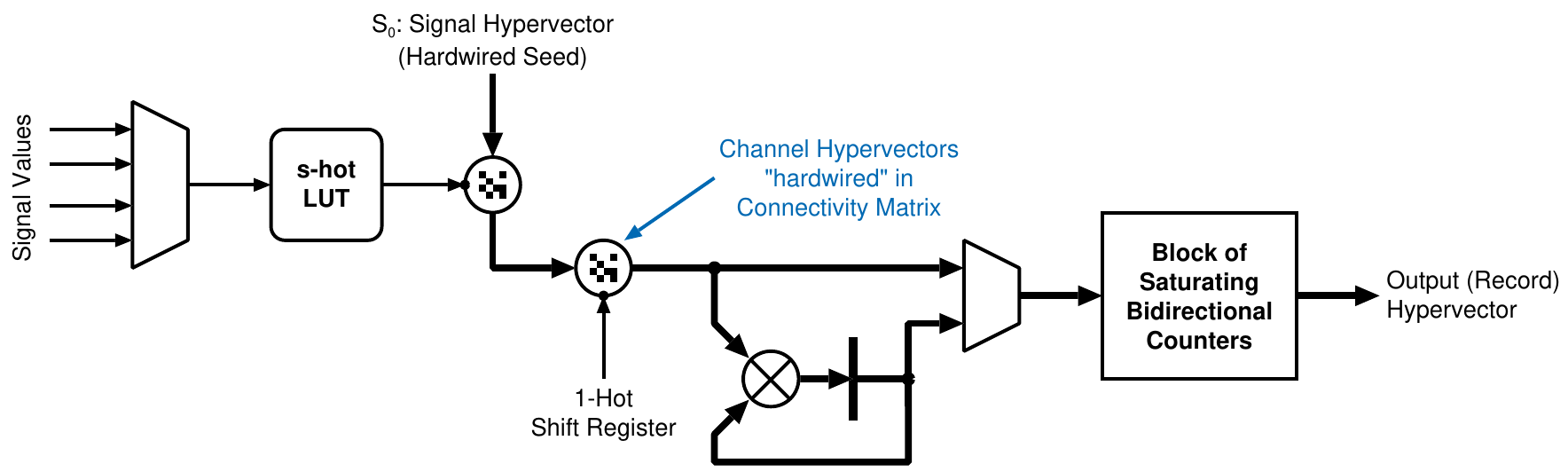}
		\captionsetup{justification=centering}
		\caption{\MAN: Sequential encoder replacing both CIM and IM with two cascaded \MAN modules.}
		\label{fig:spatial_encoder_man}
	\end{subfigure}
	\caption{The spatial encoder architectures available in the library.}
\end{figure}
The following library modules emerged from the optimizations in Section~\ref{sec:opt_mapping} and \ref{sec:spatial_encoding}:
\begin{itemize}
	\item \textit{LUT}. A purely combinational, LUT-based spatial encoder architecture. This is the starting point for optimizations and was described in \cite{EMG-HD}. See Figure~\ref{fig:spatial_encoder_lut}.
	\item \textit{CA}. A sequential spatial encoder architecture, where the IM is reproduced by a cellular automaton (CA) as described in Section~\ref{sec:opt_ca}. The bound hypervectors are bundled by a block of bidirectional saturating counters as described in Section~\ref{sec:saturation_counter}. See Figure~\ref{fig:spatial_encoder_ca}.
	\item \textit{MAN}. A sequential spatial encoder architecture, where the IM is ``hardwired'' in a manipulator's connectivity matrix as described in \ref{sec:opt_man}. The same bundling method as in the CA module is used. See Figure~\ref{fig:spatial_encoder_man}.
\end{itemize}

A summary of the CPDI of all library modules can be found in Table~\ref{tab:cpdi}.


\subsection{Temporal Encoding}
\label{sec:temp_enc}
As mentioned in Section~\ref{sec:algorithm_temporal_encoding}, the temporal encoder considers consecutive samples over time. 
This is done by rotating and binding the record hypervectors to an \textit{N-gram} hypervector:
$
\textrm{\textit{N}-gram}[t] = R[t] \otimes \rho(R[t-1]) \otimes \rho^2(R[t-2]) \otimes \cdots \otimes \rho^{N-1}(R[t-(N-1)])
$
.
%

\begin{figure}[t]
	\centering
	\includegraphics[scale=0.8]{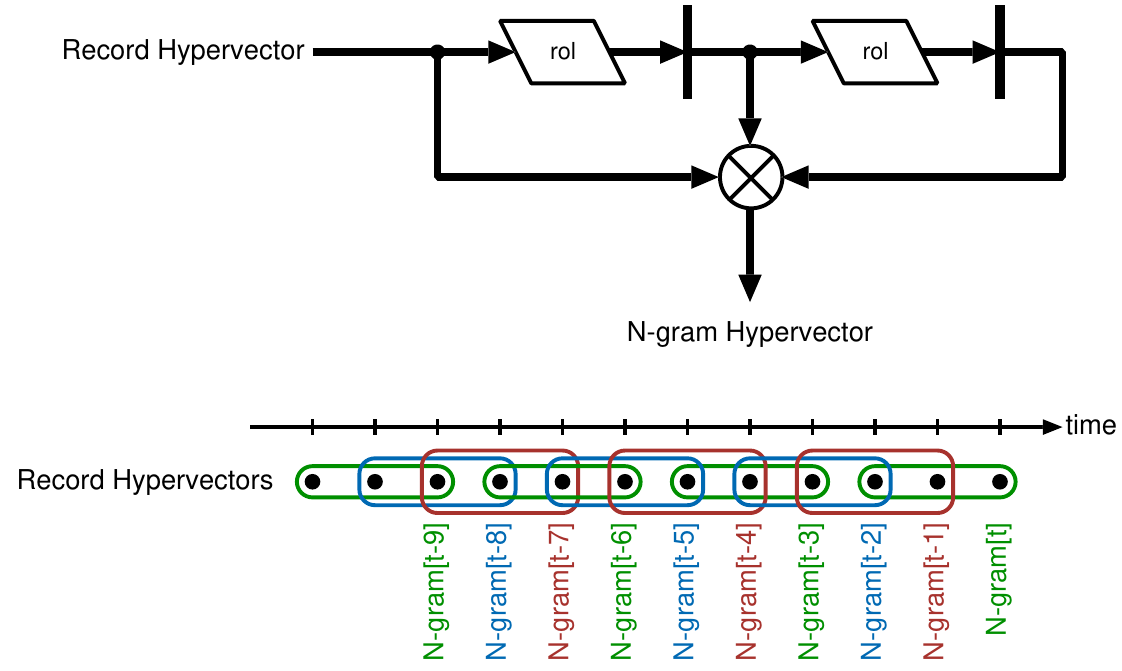}
	\caption{Top: Data dependency graph of the window-shifting $N$-gram encoder. Bottom: Depiction of the timeline of generated $N$-grams.}
	\label{ddg:window_ngram}
\end{figure}

%
In order to deliver a new $N$-gram every cycle, the records of the last $N-1$ cycles have to be kept in memory.
For this, the first record is rotated and stored. 
In the next cycle it is again rotated and stored, while the new record is rotated and stored where the last record was stored, and so on. 
In parallel, the current record is bound with all stored records and a valid $N$-gram is produced every cycle (see Figure \ref{ddg:window_ngram}).

\subsection{Bundling $N$-gram Hypervectors}
\label{sec:bundling_temporal}
All the modules that are described so far in this section form an HD projection along with a \textit{spatiotemporal} encoder.
This also constitutes a shared construct between learning and inference because the hypervectors that are produced at the output of spatiotemporal encoder (i.e., the $N$-gram hypervectors) contain all the information about the event of interest (e.g., a gesture) for training or classification. 
%
%
The AM is another part of the shared construct; however, the output of encoder queries the AM during classification while updates it during training.
For training a certain class, its $N$-gram hypervectors need to be bundled before writing into the AM. 
Different examples of a gesture are usually encoded to similar $N$-gram hypervectors, since they belong to the same class. 
This calls for a bundling method that does not require the capacity of an accurate majority function implemented with the complex saturating counters.
\subsubsection{Binarized Back-to-back Bundling as a Hardware-Friendly Approach for Approximate Bundling}
\label{sec:approx_bundling}
\begin{figure}[t!]
	\centering
	\includegraphics[width=\linewidth]{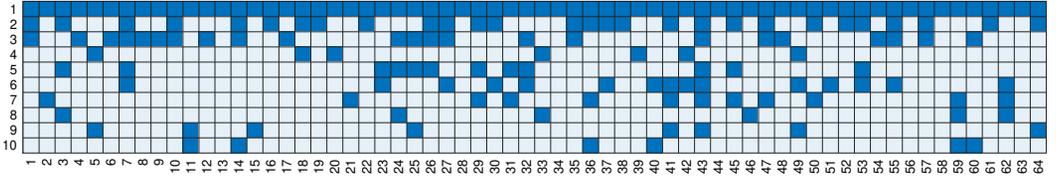}
	\caption{Example of a connectivity matrix to bundle 10 hypervectors of dimension $D=64$ iteratively.}
	\label{fig:ex_bundle_con_mat}
\end{figure}
We propose a binarized implementation of an approximate bundling operation by reusing the \MAN module.
It continuously stays in the binary space during the execution of the bundling operation, hence it enables efficient \emph{online and incremental} updates to the prototypes of the AM. 
The first step is to avoid trying to store the current majority vote and instead bundling the hypervectors iteratively, giving every vote a certain ``weight.''
This is achieved by assigning them a certain chance to be capable of turning the majority around. 
However, the vote is only turned around if the current one is different.

The first vote has a probability of $P=1$, the second $P=1/2$ and so on. 
Generally the $i$-th vote has a probability of $P_i = 1/i$ to be able to turn the majority around. 
Considering all dimensions of the hypervector, this probability turns into a weight.
In an abstract sense, these probabilities can be hardwired into the architecture with a connectivity matrix. 
For large dimensions, the $m$-th row shows $\approx D/m$ connections, which determine whether the vote at that position can turn around the majority. 
The maximum number of hypervectors in the bundling record (i.e., the rows in the connectivity matrix) should be predetermined.
Figure~\ref{fig:ex_bundle_con_mat} shows an example of connectivity matrix to bundle 10 hypervectors with dimensionality $D=64$.

\begin{figure}[t!]
	\centering
	\begin{subfigure}{0.45\textwidth}
		\centering
		\includegraphics[scale=0.75]{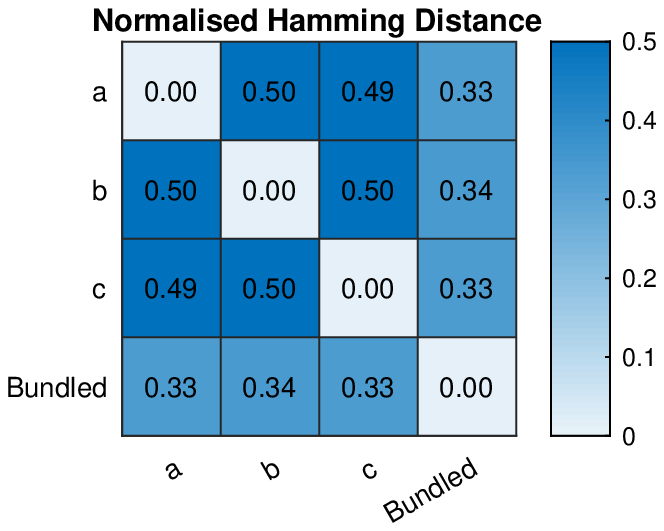}
		\caption{$a \perp b \perp c$}
	\end{subfigure}
	\hspace{0.25cm}
	\begin{subfigure}{0.45\textwidth}
		\centering
		\includegraphics[scale=0.75]{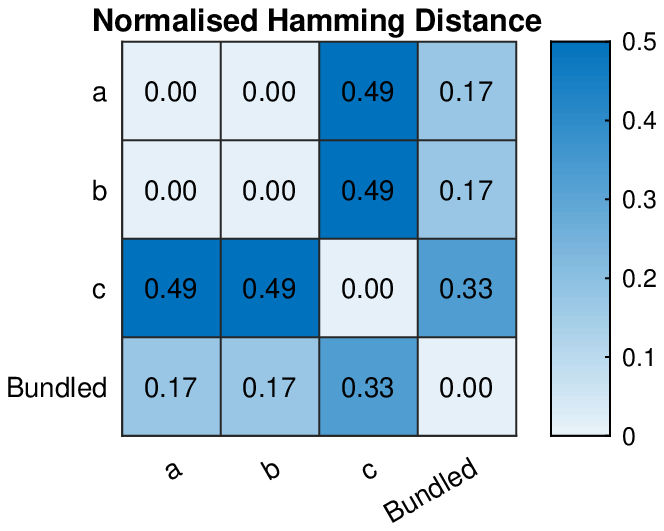}
		\caption{$a = b \perp c$}
	\end{subfigure}
	\caption{Similarity between three hypervectors and the bundled hypervector $a \oplus b \oplus c$ using binarized back-to-back bundling.}
	\label{fig:htmp_bun_3_b2b}
\end{figure}

We refer to the example of bundling three hypervectors, where two are equal and one is orthogonal. 
When bundling with the proposed approach, the orthogonal hypervector is not lost, but is similar to the record as shown in Figure \ref{fig:htmp_bun_3_b2b} (c.f. Figure~\ref{fig:htmp_bun_3_gol}). 
Furthermore, when interchanging this approximate method with the ordinary majority vote, the classification accuracy does not change.

\begin{figure}[t!]
	\centering
	\includegraphics[scale=0.8]{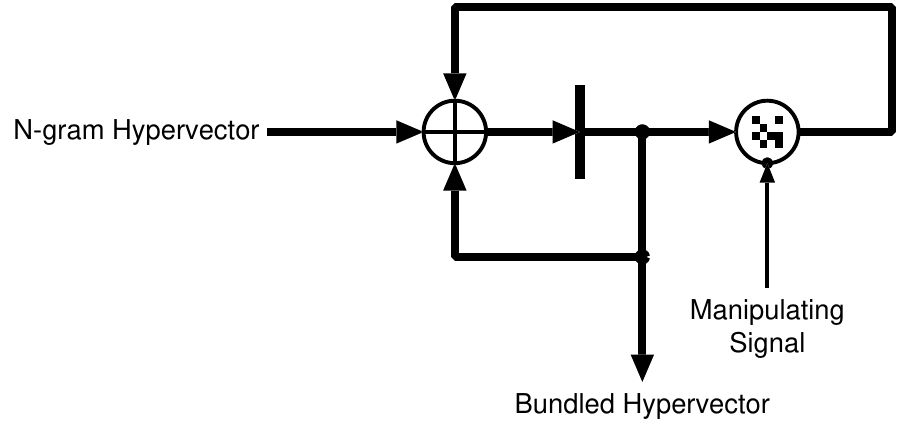}
	\caption{Data dependency graphs of the binarized back-to-back bundling to approximate temporal majority gate.}
	\label{ddg:sim_bundle}
\end{figure}

As suggested, these characteristics can be implemented using the \MAN module to generate a hypervector which is similar to the current bundled hypervector, where the Hamming distance (i.e., the degree of similarity) is determined by the connectivity matrix.
Then, the majority vote of three hypervectors is calculated from the input $N$-gram hypervector, the current bundled hypervector, and it's derived similar (manipulated) hypervector as depicted in Figure~\ref{ddg:sim_bundle}.
The similar hypervector gives the input $N$-gram hypervector a weight of $1/i$ and the current bundled hypervector a weight of $1-1/i$.
Compared to the bundling with saturating counters, this approach is far more efficient since it only requires a memory of $D$ bits (fully binarized) without adders and saturation logic.

\subsubsection{Hypervector Capacity of different Bundling Approaches} 
\label{sec:capacity}
\begin{figure}[t!]
	\centering
	\begin{subfigure}[t]{0.45\textwidth}
		\centering
		\includegraphics[width=\linewidth]{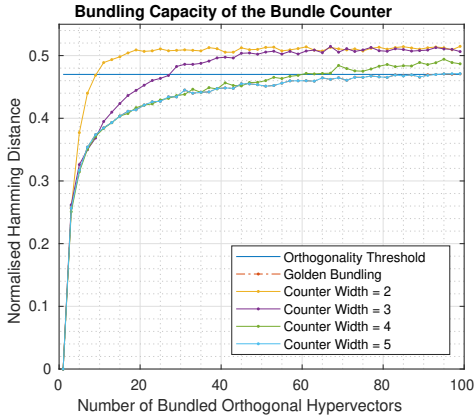}
		\caption{Capacity of the bundle counter (BC) with different widths.}
		\label{fig:capacity_bundle_counter}
	\end{subfigure}
	\hspace{0.5cm}
	\begin{subfigure}[t]{0.45\textwidth}
		\centering
		\includegraphics[width=\linewidth]{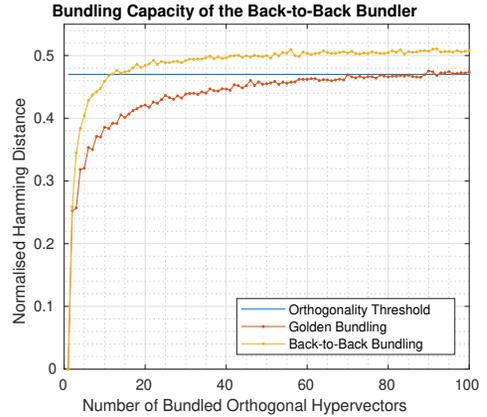}
		\caption{Capacity of back-to-back (B2B) bundling.}
		\label{fig:capacity_back-to-back}
	\end{subfigure}
	\caption{Capacity of different bundling approaches compared with the golden method for a dimensionality of $D=10000$. The graphs show the maximum normalized Hamming distance between the bundled hypervector and its compound hypervectors.}
\end{figure}
The proposed approximate bundling method slightly decreases the capacity of hypervectors.
Although for similar hypervectors, as it is the case for $N$-gram hypervectors among a class (opposed to the bound hypervectors in the spatial encoder), a large capacity is not a requirement.
Nevertheless, it is necessary to evaluate how much information a hypervector can store, or how many hypervectors can be bundled into a hypervector (i.e., the capacity of a bundling method).

The capacity can be measured by bundling an increasing number of orthogonal hypervectors and trying to recall the information by measuring the similarity between the bundled hypervector and all compound hypervectors. 
As long as none of the compound hypervectors crosses the orthogonality threshold (see Section \ref{sec:algorithm_orthogonality}), their information is still contained in the bundled hypervector. 
As soon as one of the compound hypervectors becomes orthogonal to the bundled, the bundling method has failed to capture all the information.

For comparison, the ordinary majority vote (see Section \ref{sec:algorithm_bundling}) is used as the reference bundling method. 
This approach is referred to as the \textit{golden} method. 
The two other approaches are the binarized \textit{back-to-back (B2B)} method from Section \ref{sec:approx_bundling} and the \textit{bundle counter (BC)} method (Section \ref{sec:saturation_counter}), which can be viewed as a very close approximation of the golden method.

The capacity of the binarized back-to-back method in comparison with the golden method is depicted in Figure \ref{fig:capacity_back-to-back}.
The golden method is capable of storing the information of about 60--70 orthogonal hypervectors for a dimensionality of $D=10000$, whereas the back-to-back binary method saturates between 10--15 hypervectors.

However, the capacity of the counter method is dependent on the number of bits (i.e., width) used to represent the current vote. 
The smaller the width, the fewer the resources required but the smaller its capacity. 
This can be seen in Figure \ref{fig:capacity_bundle_counter}. 
We observe that a width of 5 bits is sufficient to achieve the same capacity as the golden method. 
When bundling fewer hypervectors, the width should be adjusted to ones needs to minimize the required resources.

%
%


\subsection{Library: Temporal Encoder Modules}

\begin{figure}[t!]
	\centering
	\begin{subfigure}{\textwidth}
		\centering
		\includegraphics[scale=0.7]{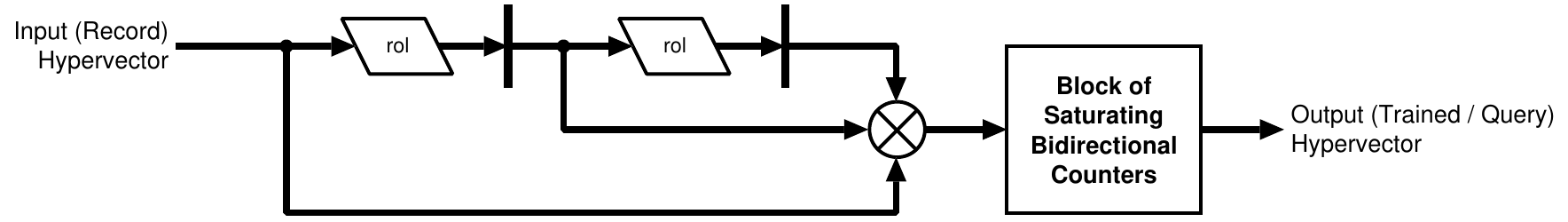}
		\caption{Bundle counter (BC): Using saturating bidirectional counters to bundle $N$-gram hypervectors.}
		\label{fig:temporal_encoder_bc}
	\end{subfigure}
	\vspace{0.5cm}
	
	\begin{subfigure}{\textwidth}
		\centering
		\includegraphics[scale=0.7]{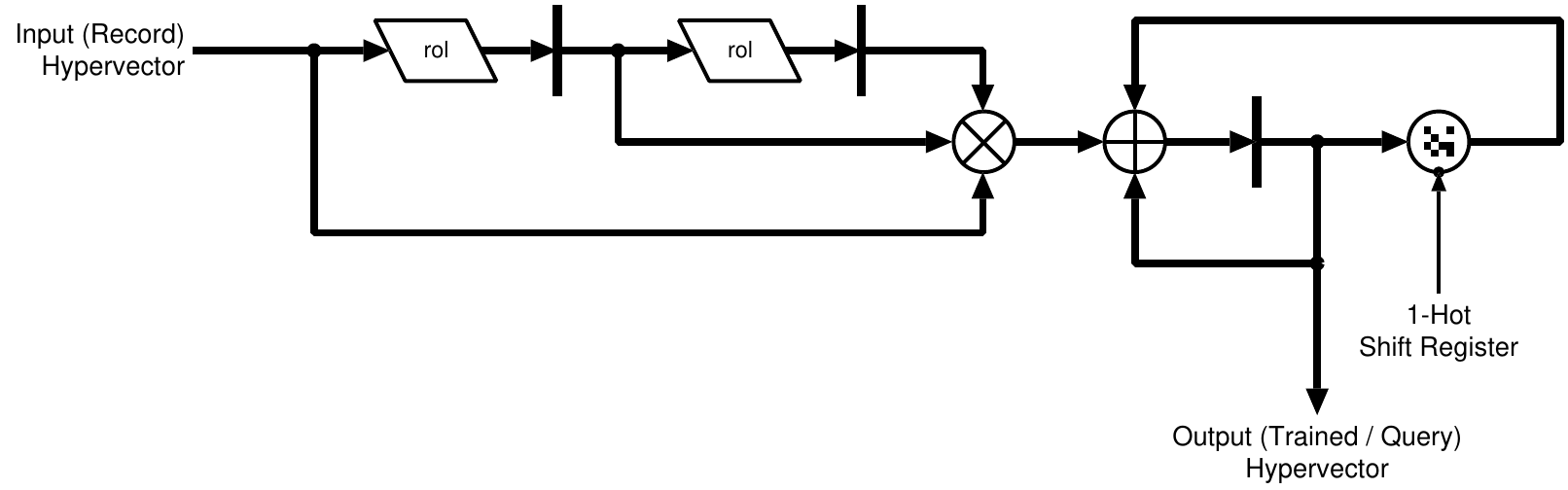}
		\caption{Binarized Back-to-back (B2B): Using the \MAN module to approximate bundling of $N$-gram hypervectors.}
		\label{fig:temporal_encoder_b2b}
	\end{subfigure}
	\caption{The temporal encoder architectures available in the library.}
\end{figure}

The following library modules emerged from the optimizations in Section~\ref{sec:temp_enc} and \ref{sec:bundling_temporal}:
\begin{itemize}
	\item \textit{BC}. A temporal encoder architecture using counter-based bundling as described in Section~\ref{sec:temp_enc} and \ref{sec:saturation_counter}. See Figure~\ref{fig:temporal_encoder_bc}.
	\item \textit{B2B}. A temporal encoder architecture using manipulator-based back-to-back binary bundling as described in Section~\ref{sec:temp_enc} and \ref{sec:approx_bundling}. See Figure~\ref{fig:temporal_encoder_b2b}.
\end{itemize}

\subsection{Associative Memory (AM)}
\label{sec:associative_memory}
The associative memory (AM) is the part of the architecture that is the most challenging to optimize.
One reason is the memory required to store the ``trained'', prototype, or rather the bundled hypervectors that represent the classes. 
Another reason is the nature of the Hamming distance, that has to be computed between the query hypervector---of which we want to find the class it belongs to---and each trained hypervectors.

As described in Section \ref{sec:algorith_hamming}, the Hamming distance measures the number of positions at which two hypervectors differ. 
This is equal to computing the population count of a hypervector binding those two hypervectors. 
So far, digital methods for AMs count through all components resulting in a classification latency in the order $\mathcal{O}(D)$~\cite{Rahimi2016, TCAS17,HAM_HPCA17,VoiceHD_ICRC17}.
We focus on reducing this latency by adding up all hypervector components.

\subsubsection{Deep Adder Trees} \label{sec:deep_ad_trees}
When trying to add up all bits of a hypervector, working with tree structures is the most efficient way.
In this manner, the AM takes only one clock cycle to compute the Hamming distance, at a cost to long logic delay and gate counts.
For a perfect binary tree, which is the case for hypervectors of dimension $D=2^n$, the depth is 
$\log_2(D) = n$ which is also the number of adder stages. 
The amount of adders in stage $i$ is
$D/2^i$ and the width of the adders in stage $i$ equals to $i$. 
In the simple case of using ripple-carry-adders, the logic delay of the adder tree is equal to
$
\sum_{i=1}^n i = \frac{n(n+1)}{2}
$
delays of a 1-bit-adder.
For a dimension $D=2^{13}=8192$, this amounts to the delay of $91$ 1-bit-adders, which will most likely result in the longest path in the architecture. This could be reduced with pipeline registers close to the root, i.e., the final result. 
The total equivalent of 1-bit-adders for the whole tree can be calculated as follows:
$
\sum_{i=1}^n \frac{D \cdot i}{2^i}
$
which, for a dimension $D=2^{13}$ yields 16369 1-bit-adders.

Although this number of adders seems very high, an FPGA can handle it easily with lookup tables. 
Furthermore, using the counters as an alternative might seem resource friendlier at first, but turns out an incompetent choice. 
The reason is that each bit of the hypervector somehow has to be directed to the counter. 
This requires either huge multiplexers or shift registers with input multiplexers, which both leads to immense area overhead. 
While the overhead is considerable, the cycles needed to compute the Hamming distance is of the order $\mathcal{O}(D)$. 
This is a poor trade-off compared to the high throughput and moderate overhead of adder tree architectures.

Using the adder trees to compute the Hamming distance between two hypervectors, two AM variations emerge. 
A fully parallel architecture with replicated adders, leading to $\mathcal{O}(1)$ computation cycles, and a vector-sequential architecture, which shares one adder tree to compute the Hamming distance of all hypervectors one after the other, hence leading to $\mathcal{O}(n_\textrm{classes})$ computation cycles.

\subsection{Library: Associative Memory Modules}

\begin{figure}[h]
	\centering
	\begin{subfigure}{\textwidth}
		\centering
		\includegraphics[scale=0.7]{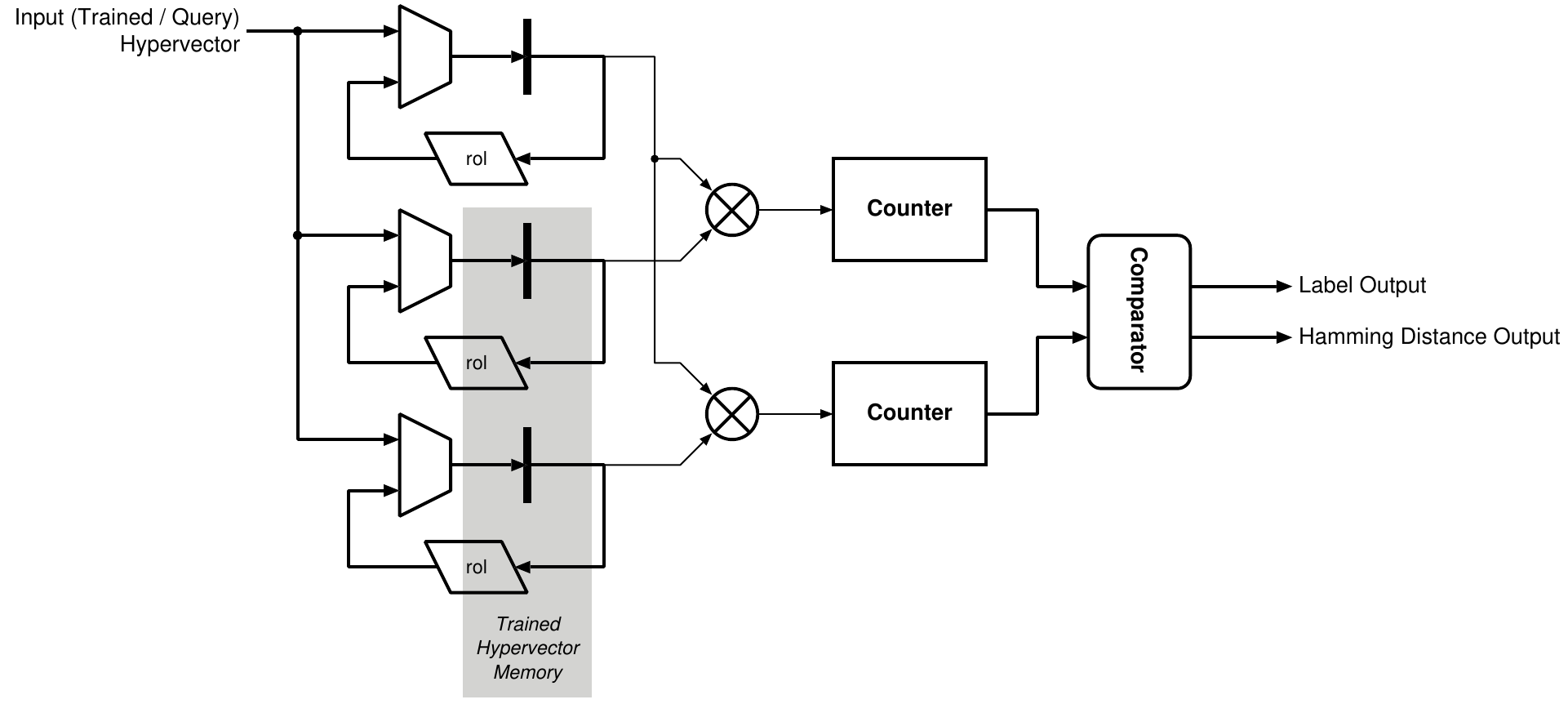}
		\caption{Bit-sequential (BS): This AM has a latency of $\mathcal{O}(D)$.}
		\label{fig:associative_memory_bs}
	\end{subfigure}
	\vspace{0.5cm}
	
	\begin{subfigure}{\textwidth}
		\centering
		\includegraphics[scale=0.7]{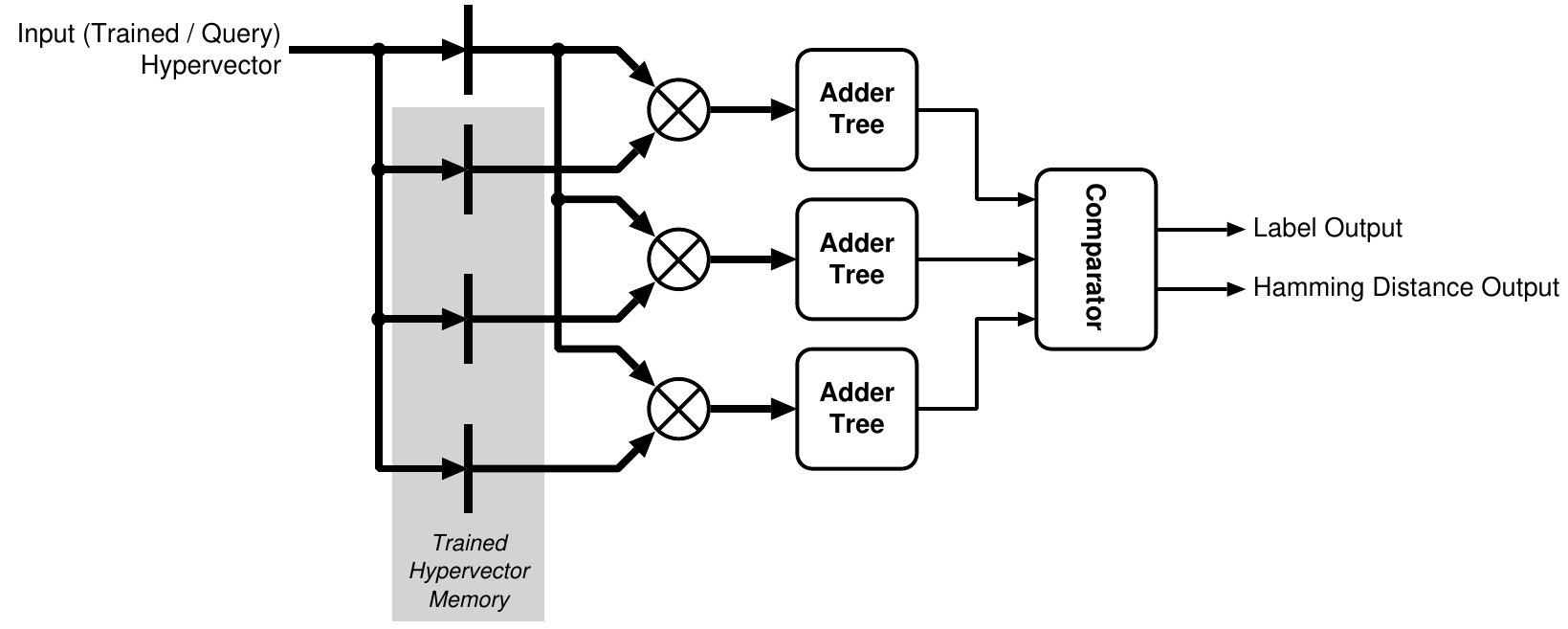}
		\caption{Combinational (CMB): This single-cycle AM has a latency of $\mathcal{O}(1)$.}
		\label{fig:associative_memory_cmb}
	\end{subfigure}
	\vspace{0.5cm}
	
	\begin{subfigure}{\textwidth}
		\centering
		\includegraphics[scale=0.7]{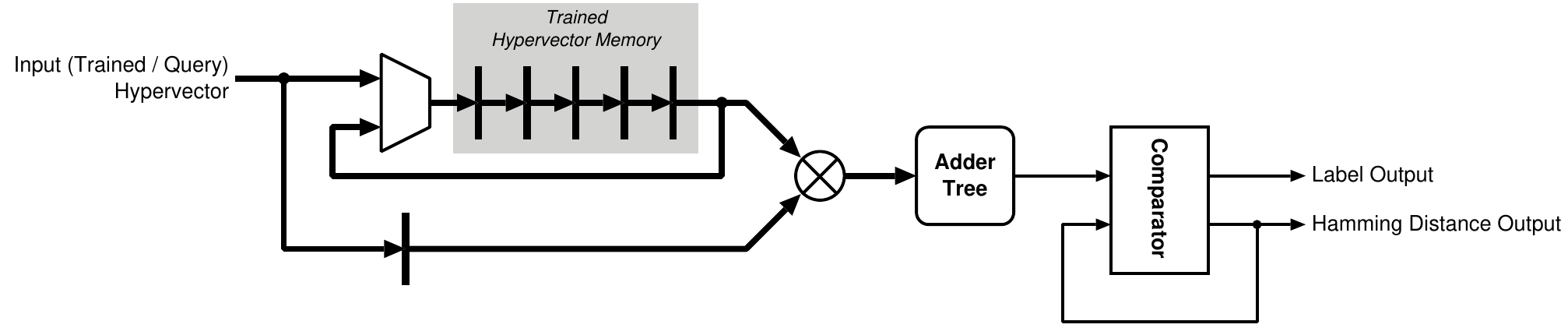}
		\caption{Vector-sequential (VS): This AM has a latency of $\mathcal{O}(n_\textrm{classes})$.}
		\label{fig:associative_memory_vs}
	\end{subfigure}
	\caption{The associative memory architectures available in the library.}
\end{figure}

The following library modules emerged from the optimizations in Section~\ref{sec:associative_memory}:
\begin{itemize}
	\item \textit{BS}. A bit-sequential AM architecture. This is the starting point for optimizations and was described in~\cite{Rahimi2016, TCAS17,HAM_HPCA17,VoiceHD_ICRC17}. See Figure~\ref{fig:associative_memory_bs}.
	\item \textit{CMB}. An AM architecture based on adder trees as described in Section~\ref{sec:deep_ad_trees}. See Figure~\ref{fig:associative_memory_cmb}.
	\item \textit{VS}. A vector-sequential AM architecture based on adder trees as described in Section~\ref{sec:deep_ad_trees}. See Figure~\ref{fig:associative_memory_vs}.
\end{itemize}

\section{Design Space Exploration and Experimental Results}
\label{sec:design_space}
In order to evaluate the library modules, they are configured for the EMG-based hand gesture recognition task, and all possible combinations of HD architectures (i.e., our design space) are synthesized for a Xilinx\textsuperscript{\textregistered} Virtex UltraScale\textsuperscript{\texttrademark} FPGA \cite{ds_xilinx_us}.
All the HD architectures are functionally equivalent and exhibit iso-accuracy.
The parameters for the configured architectures are listed in Table~\ref{tab:synth_configs}.
The library can be configured to conduct virtually any learning and classification task.

\begin{table}[t]
	\centering
	\caption{Parameter configuration for the case study.}
	\label{tab:synth_configs}
	\begin{tabular}{>\raggedright m{0.4\textwidth}M{0.175\textwidth}}
		\toprule
		\textbf{Parameter}						& \textbf{Value} \\
		\hline
		Hypervector Dimension ($D$)					& 8192 \\
		Channels								& 4 \\
		Classes									& 5 \\
		Quantization							& 21 \\
		$N$-gram Size							& 3 \\
		Bundle Counter Width (\textit{MAN} \& \textit{CA})			& 3 \\
		Bundle Counter Width	 (\textit{BC})				& 5 \\
		Max. Bundle Cycles (\textit{B2B})					& 256 \\
		\bottomrule
	\end{tabular}
\end{table}

\begin{figure}[t]
	\centering
	\includegraphics[scale=0.7]{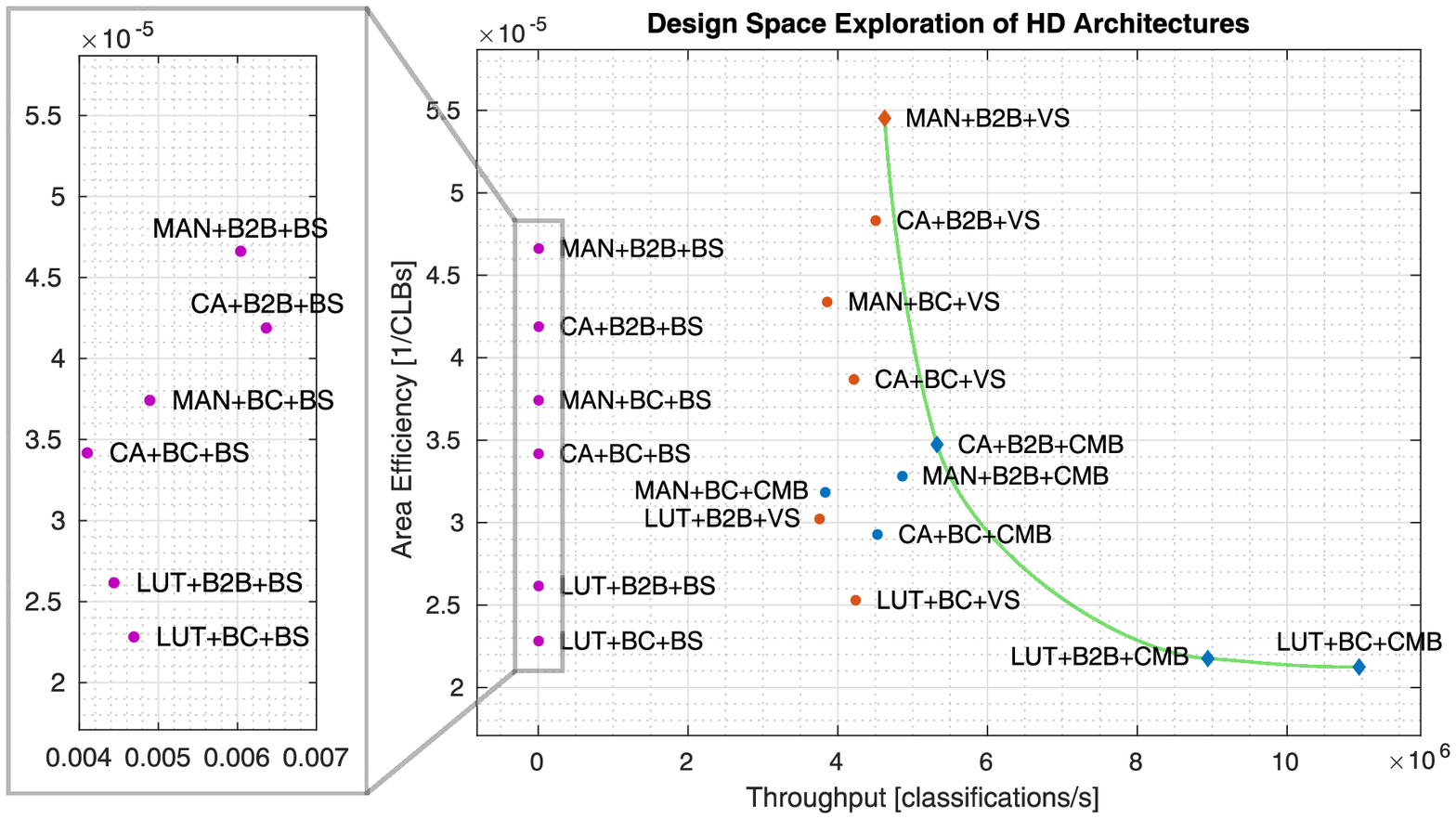}
	\caption{Design space exploration of HD architectures using all possible combinations of the modules available in the library. Colors indicate the architectures with the same type of AM. Pareto optimal architectures are marked with a diamond $\Diamondblack$ and connected by a green line representing the Pareto frontier.}
	\label{fig:design_space}
\end{figure}

Each HD architecture is composed of three modules in series: a type of mapping and spatial encoder followed by a type of temporal encoder, and finally a type of AM.
To conduct the design space exploration, each architecture's throughput is plotted against its area efficiency (defined as $1/\textrm{CLBs}$) in Figure~\ref{fig:design_space}.
The quality of an architecture increases when going from left to right and/or bottom to top.
The color coding represents HD architectures with the same type of AM.
%

Our starting point is the LUT+BC+BS architecture as an improved version of ~\cite{EMG-HD} using bidirectional saturating counters.
What can be observed is that by replacing the LUT module with the proposed MAN and CA modules, a significant area saving is achieved.
This area saving is consistent with any combination of temporal encoder and AM.
A similar area improvement can be observed when replacing the BC module with the novel B2B module. 
Combining both optimization leads to an area improvement of up to $\times 2.39$.
On the other hand, a massive throughput improvement of up to $\times 2337$ can be achieved by moving from an AM with the BS module to VS and finally CMB.

\begin{table}[t]
	\centering
	\caption{Area and throughput results of the ``starting point'' and the Pareto optimal architectures.}
	\label{tab:design_space}
	\begin{tabular}{lM{0.18\linewidth}M{0.18\linewidth}M{0.18\linewidth}M{0.18\linewidth}}
		\toprule
		\textbf{Architecture} & \textbf{Throughput [$\textrm{classifications}/s$]} & \textbf{Throughput Improvement} & \textbf{Area [$\textrm{CLBs}]$} & \textbf{Area Improvement} \\
		\hline
		LUT+BC+BS 		& $4.69\cdot10^3$	& $\times 1$ 	& $43825$	& $\times 1$ \\
		\hline
		MAN+B2B+VS 		& $4.62\cdot10^6$ 	& $\times 986$	& $18340$	& $\times 2.39$ \\
		CA+B2B+CMB		& $5.33\cdot10^6$	& $\times 1136$	& $28788$	& $\times 1.52$ \\
		LUT+B2B+CMB		& $8.94\cdot10^6$	& $\times 1906$	& $45961$	& $\times 0.95$ \\
		LUT+BC+CMB		& $10.96\cdot10^6$	& $\times 2337$ & $47068$	& $\times 0.93$ \\
		\bottomrule
	\end{tabular}
\end{table}

Different combinations of the modules produce architectures with varying area/throughput improvements.
Eventually, four architectures stand out as pareto optimal architectures (see Table~\ref{tab:design_space}).
These offer different trade-offs and can be selected depending on the user's requirements.
The throughput of these architectures is significantly higher than the classification constraint for real-time EMG tasks~\cite{EMG_RealTime_10ms,EMG_RealTime_10ms_2}. 
Note that different configurations may lead to different pareto optimal architectures.

\subsection{Scalability: Larger number of Channels and Classes}
Here, we assess the scalability of our proposed methods when doubling the number of channels and classes.
The spatial encoder with the CA module shows the best area efficiency for applications with a larger number of channels, followed by the spatial encoder with the \MAN module.
The memory footprint of CA module is independent of the number of channels since only a seed hypervector to initialize the CA state needs to be stored, hence the area will not increase (see Table~\ref{tab:spatial_encoder_scaling}).
However, it requires almost twice clock cycles to produce the channel hypervectors for the doubled number of channels.
The spatial encoder with the LUT shows opposite scalability: it maintains almost the same throughput but increases the area by $2.41 \times$.
Focusing on the AM module, an application with twice the number of classes will impose a larger area to the CMB and BS modules, whereas the VS' area is mostly unaffected, apart from the storage for additional trained hypervectors (see Table~\ref{tab:associative_memory_scaling}).
%

\begin{table}[t!]
	\centering
	\caption{Scalability of the library modules.}
	\label{tab:module_scaling}
	\begin{subtable}[t]{0.45\textwidth}
	    \centering
    	\caption{Throughput and area scaling of the spatial encoder modules when doubling the number of channels from $4$ to $8$.}
    	\label{tab:spatial_encoder_scaling}
    	\begin{tabular}{lM{0.3\linewidth}M{0.3\linewidth}}
    		\toprule
    		\textbf{Module} & \textbf{Throughput Scaling} & \textbf{Area Scaling} \\
    		\hline
                LUT    & $\times 0.94$               & $\times 2.41$          \\
                CA     & $\times 0.45$               & $\times 0.99$          \\
                MAN    & $\times 0.61$               & $\times 1.01$          \\
    		\bottomrule
    	\end{tabular}
	\end{subtable}
	\hspace{0.3cm}
	\begin{subtable}[t]{0.45\textwidth}
	    \centering
    	\caption{Throughput and area scaling of the AM modules when doubling the number of classes from $6$ to $12$.}
    	\label{tab:associative_memory_scaling}
    	\begin{tabular}{lM{0.3\linewidth}M{0.3\linewidth}}
    		\toprule
    		\textbf{Module} & \textbf{Throughput Scaling} & \textbf{Area Scaling} \\
    		\hline
                BS      & $\times 0.49$               & $\times 1.89$          \\
                CMB     & $\times 0.63$               & $\times 2.14$          \\
                VS      & $\times 0.59$               & $\times 1.10$          \\
    		\bottomrule
    	\end{tabular}
	\end{subtable}
\end{table}


\section{Conclusions}
This paper proposes hardware optimizations---in an open-source VHDL library---for dense binary HD computing that enable efficient synthesis of acceleration engines handling both inference and training tasks on an FPGA.
The Pareto optimal design is mapped on only $18340$ CLBs of a Xilinx\textsuperscript{\textregistered} UltraScale\textsuperscript{\texttrademark} FPGA achieving simultaneous $2.39\times$ lower area and $986\times$ higher throughput compared to the baseline.
This is accomplished by:
(1) rematerializing hypervectors on the fly by substituting the cheap logical operations for the expensive memory accesses to seed hypervectors;
(2) online and incremental learning from different gesture examples while staying in the binary space;
(3) combinational associative memories to steadily reduce the latency of classification.  
%
%
Our future work will target an ASIC implementation of the library modules.

\begin{acks}
Support was received from the ETH Zurich Postdoctoral Fellowship program, the Marie Curie Actions for People COFUND Program, and the European Union's Horizon 2020 Research and Innovation Program through the project MNEMOSENE under Grant 780215.
\end{acks}


\bibliographystyle{ACM-Reference-Format}
\bibliography{main}

\end{document}